\documentclass[%
 aip,
 amsmath,amssymb,
 reprint,%
nofootinbib
]{revtex4-1}

\usepackage{graphicx}
\usepackage{dcolumn}
\usepackage{bm}

\usepackage[utf8]{inputenc}
\usepackage[T1]{fontenc}
\usepackage{mathptmx}
\usepackage{etoolbox}
\usepackage{siunitx}
\usepackage{booktabs}
\usepackage{multirow}
\usepackage{xcolor}
\usepackage{graphicx}
\usepackage{hyperref}

\graphicspath{{figures/}}

\DeclareMathAlphabet{\mathcal}{OMS}{cmsy}{m}{n}

\makeatletter
\let\NAT@citesuper\NAT@cite
\makeatother

\setcitestyle{numbers,square,comma,sort&compress}

\newcommand{\eqnref}[1]{Eq.~(\ref{#1})}
\newcommand{\figref}[1]{Fig.~\ref{#1}}
\newcommand{\tabref}[1]{Table~\ref{#1}}

\begin{document}

\preprint{AIP/123-QED}

\title[\textit{DoS Dos and Don'ts}]{DoS Dos and Don'ts}

\author{Lucas Warwaruk}
\affiliation{Department of Mechanical Engineering, Massachusetts Institute of Technology, Cambridge, Massachusetts 02139.}

\author{Konstantinos Zinelis}
\affiliation{Department of Chemical Engineering, Massachusetts Institute of Technology, Cambridge, Massachusetts 02142.}

\author{Randy H. Ewoldt}
\affiliation{Department of Mechanical Science and Engineering, University of Illinois Urbana-Champaign, Urbana, Illinois 61801.}

\author{Christopher W. Macosko}
\affiliation{Department of Chemical Engineering and Materials Science, University of Minnesota, Minneapolis, Minnesota 55455.}

\author{Gareth H. McKinley}
\homepage{Author to whom correspondence should be addressed; electronic mail: gareth@mit.edu}
\affiliation{Department of Mechanical Engineering, Massachusetts Institute of Technology, Cambridge, Massachusetts 02139.}

\date{\today}

\begin{abstract}
Dripping-onto-Substrate (DoS) rheometry is a well-established method for measuring the extensional rheology of low-viscosity liquids. However, clear guidelines on the capabilities and limitations of the technique are lacking. In the present work, we define operational limits for measuring a transient extensional viscosity directly from observation of the rate of filament thinning, as well as model-based bounds on calculating a viscosity $\eta$ and extensional relaxation time $\tau_E$ of a liquid using DoS. Dilute solutions of polyethylene oxide (PEO) and polyacrylamide (PAM) are used to probe the lower limit of measurable $\tau_E$, demonstrating that values as low as 0.1 ms can be resolved, provided (a) the intrinsic Deborah number (based on the ratio of the relaxation time and the Rayleigh breakup time scale) is $De \geq \mathcal{O}(0.1)$ and (b) an instrumental constraint related to spatial and temporal resolution is satisfied. This instrumental constraint is quantified through a new metric we define as the \textit{filament capture rate}, a ``figure of merit'' (expressed in Hz) that can be used to quantify the number of data points within the elasto-capillary regime that are available for extraction of $\tau_E$. We also investigate the sensitivity to other experimental parameters including variations in nozzle radius and Bond number ($Bo$). Across the tested range ($0.2 < Bo < 0.7$), extensional relaxation times for the same fluid vary by less than $\pm16$ \%; however, experiments with low viscosity fluids at $Bo > 0.5$ exhibit damped gravitational oscillations that affect early-time dynamics. Collectively, these results provide a quantitative roadmap for reliable DoS rheometry and affirm its use for measuring sub-millisecond relaxation times in weakly elastic fluids.
\end{abstract}

\maketitle

\section{Introduction}\label{sec:1}

Adding a small amount of high molecular weight polymer to a Newtonian liquid can have an appreciable influence on the rheology of the resulting solution. In particular, dilute solutions of long chain polymers can exhibit an exceptionally large resistance to extensional deformation, despite having a shear viscosity comparable to the Newtonian solvent. The large extensional viscosity of dilute polymer solutions plays a critical role in a number of practical applications, for example, preventing mist formation of flammable hydrocarbon fuels \citep{Lhota2024}, or reducing drag in turbulent pipe flows \citep{Lumley1969,Virk1975,Rajappan2019}. A number of novel experimental techniques exist for measuring the extensional features of low viscosity non-Newtonian fluids, and have been reviewed in detail by \citet{Galindo2013}, \citet{Haward2016}, \citet{delGuidice2017} and \citet{delGuidice2022}. 

For such materials, extensional rheometry techniques that rely on capillary forces to drive a filament thinning flow have become a preferred choice \citep{Rodd2005,Campo2010,Keshavarz2016,Dinic2015,Rajesh2022}. Viewed broadly, such techniques can be thought of as a quantitative version of the protorheological ``thumb-and-forefinger test'' \citep{Mckinley2005,Hossain2024}. One such technique, first demonstrated by \citet{Bazilevsky2011}, then developed by \citet{Dinic2015} is known as Dripping-onto-Substrate (DoS) extensional rheology. This has proven to be an effective and relatively simple-to-implement method for measuring the extensional characteristics of dilute polymer solutions \citep{Sur2018,Dinic2019,Robertson2022,Soetrisno2023}. 

In DoS rheometry, a small drop of liquid is formed at the tip of a blunt-end nozzle and the pendant drop  is slowly brought into contact with a hydrophilic substrate. Computational simulations demonstrate that the energy released by the pendant droplet spreading across a partially-wetting substrate drives a rapid initial neck formation \citep{Zinelis2024}. Capillarity subsequently produces a strong transient extensional deformation in the thinning fluid ligament, especially in low viscosity fluids. High-speed digital imaging can be used to monitor the evolution in the filament radius $R_{\mathrm{min}}(t)$ over time, from which the time-to-break $t_b$ and extensional relaxation time $\tau_E$ can be assessed \citep{Dinic2015,Ng2021}. If the liquid-vapour surface tension is known, the transient extensional viscosity $\eta_E^+(t)$, and the dynamic viscosity $\eta$ (for Newtonian liquids) can be extracted \citep{Mckinley2000,Anna2001}. 

Since its conception, numerous investigations have relied on DoS rheometry to measure the extensional characteristics of a variety of low viscosity fluids \citep{Dinic2017,Rosello2019,Narvaez2021,Lauser2021,Robertson2022,Zhang2022,Zhang2024}. However, a clear set of experimental guidelines on the capabilities and limitations of DoS rheometry is lacking. Therefore, the goal of the current investigation, is to establish the key ``Dos and Don'ts'' of DoS rheometry; discussing critical parameters that influence the quality of the measurements and the accuracy of the resulting estimates of the extensional relaxation time. However, DoS rheometry is not the only capillarity-driven method for measuring the extensional relaxation time $\tau_E$, and this introduction will briefly summarize some of the other capillarity-driven approaches for measuring extensional rheology. 

The capillarity-driven thinning method based on a rapid initial plate separation (as implemented in the Capillary Break-up Extensional Rheometer or CaBER instrument) has been widely used for measuring a transient extensional viscosity and relaxation time of various dilute and semi-dilute polymer solutions \citep{Anna2001,Rodd2005,Clasen2006a}. In the CaBER device a liquid sample is first loaded between two circular plates to form a cylindrical liquid bridge. An axial step strain is then imposed on the sample by rapidly separating the plates apart from one another. Capillary forces drive the thinning, and ultimate break-up, of the unstable liquid bridge, producing a transient uniaxial extensional flow from which the time-to-break $t_b$, transient extensional viscosity $\eta_E^+$, and extensional relaxation time $\tau_E$ can be extracted. \citet{Rodd2005} used scaling arguments to establish an ``operability diagram'' for CaBER, demonstrating that the technique can be used to measure relaxation times as small as $\tau_E \approx 1$ ms for low viscosity ($\eta < 70$ mPa s), aqueous polyethylene oxide (PEO) solutions.

Since the work of \citet{Rodd2005}, a number of experimental works have developed new or modified capillarity-driven methods that are capable of measuring relaxation times less than 1 ms. For example, \citet{Campo2010} measured relaxation times down to $\tau_E = 240$ \SI{}{\micro\second} using a modified CaBER approach known as the Slow Retraction Method (SRM) in which the plates are slowly separated at a constant velocity until the elongated liquid bridge crosses the static stability limit and rapidly fails. \citet{Keshavarz2015} measured relaxation times as small as $\tau_E =  60$ \SI{}{\micro\second} using a jet-based technique they christened Rayleigh Ohnesorge Jetting Extensional Rheometry (ROJER). The ROJER method, like DoS and CABER, relies on the Rayleigh-Plateau instability to drive filament thinning, but in an axially-perturbed fluid jet traveling at a high velocity. Subsequent to the work of \citet{Keshavarz2015}, an updated and more cost-effective ROJER design was developed by \citet{Greiciunas2017}. The recent development and growth of DoS rheometry has also produced a number of experimental investigations that have measured extensional relaxation times on the order of 10 \SI{}{\micro\second} $\leq \tau_E \leq$ 100 \SI{}{\micro\second} \citep{Dinic2015,Sur2018,Rosello2019,Zhang2024}. The ability to make measurements of very small relaxation times in DoS, SRM and ROJER, is facilitated by the lack of fluid inertia-related issues that are common in the rapid step-strain CaBER protocol when testing mobile liquids such as polymer solutions that are characterized by low viscosities and $\tau_E < 1$ ms. 

A number of investigations have compared the filament thinning dynamics and resulting values of extensional relaxation time for the same fluid measured using the different methods of CaBER, SRM, ROJER, and DoS \citep{Mathues2018,Gaillard2024}. Recently, \citet{Gaillard2024} found that the relaxation times obtained from analysis of CaBER, SRM, and DoS measurements can be quite different, and also demonstrated a strong dependence on test geometry (i.e., the plate or nozzle diameter used in CaBER or DoS respectively). \citet{Aisling2024} argued that disagreements between the relaxation time measured using different methods (e.g., CaBER and DoS) can be attributed to their specific kinematic history, and that reliable measurements of relaxation time are predicated on the establishment of a sufficiently large extensional strain rate, i.e., $\dot{\varepsilon} > 2/(3\tau_E)$, in the initial step strain period that forms the elongated liquid bridge. A detailed recent study by \citet{Allgood2025} with Newtonian calibration oils has shown that the thinning dynamics and resulting measurements of transient extensional viscosity are controlled by careful selection of an appropriate aspect ratio characterizing the ratio between the end-plate diameter and the axial separation distance used in CaBER. Despite these notable recent caveats regarding the interpretation of filament thinning rheometry, a number of studies have shown that all of these capillarity-driven methods (i.e., CaBER, SRM, ROJER and DoS) can provide reliable estimates of the extensional relaxation time and transient extensional viscosity provided sufficient care is taken to carefully select the appropriate experimental conditions \citep{Campo2010,Vadillo2012,Keshavarz2016,Soetrisno2023}.

Of the different filament thinning rheometers that rely on capillarity as a driving force, DoS is the least complex to implement. While CaBER, SRM, and ROJER require specialized hardware such as a precision axial displacement stage or piezoelectric actuators, a typical DoS setup can be constructed from standard components that can be found readily in most rheology laboratories. These components include a syringe pump, small gauge blunt-end nozzle, a bright diffuse light source and a high-speed camera. However, a detailed guide on imaging magnification, frame rate and appropriate operating spaces is missing in the literature. Establishing best practices for DoS rheometry will help further popularize its use, making extensional rheology measurements of low viscosity fluids increasingly accessible to a wider audience.

In the present work we use physical considerations, including scaling arguments, to establish the parameter range over which the extensional viscosity can be measured, as well as the critical Deborah $De$ and Ohnesorge $Oh$ number, below which the relaxation time $\tau_E$ and dynamic viscosity $\eta$ cannot be reliably measured with DoS. Dilute solutions of a 2 MDa and an 8 MDa PEO, as well as a highly polydisperse drag-reducing polyacrylamide (PAM) polymer with a (manufacturer-reported) viscosity average molecular weight of $M_v \approx 20$ MDa are used as test fluids for probing the critical Deborah number $De$, and other experimental control parameters. Recent numerical simulations of DoS by \citet{Zinelis2024} using the Oldroyd-B and FENE-P (finitely extensible non-linear elastic dumbbell model with the Peterlin closure approximation) constitutive equations have systematically explored the influence of gravity, substrate wettability, elasticity and polymer finite extensibility on the filament thinning dynamics. The same range of parameters probed numerically by \citet{Zinelis2024} are explored experimentally in the current work.

\section{Operating limits of capillarity-driven extensional rheometry}\label{sec:2}

In the following section, we provide a general discussion of operating limits for capillarity-driven extensional rheometry techniques, such as DoS. The dynamics of filament thinning for Newtonian and viscoelastic fluids are first reviewed in \S\S \ref{sec:2.A} and \ref{sec:2.B}, respectively. In these subsections, commonly-used analytical expressions for the minimum measurable viscosity $\eta$ (for Newtonian fluids) and extensional relaxation time $\tau_E$ (for viscoelastic fluids) are derived. In \S \ref{sec:2.C} the experimental limitations of DoS rheometer hardware are detailed. A new figure of merit we define as the \textit{filament capture rate} is established. This metric can be used to compare and contrast different experimental embodiments and evaluate if DoS measurements are sufficiently resolved to accurately extract values of the extensional relaxation time $\tau_E$. The operating limits derived in \S \ref{sec:2.A} pertain to Newtonian liquids, while those detailed in \S\S \ref{sec:2.B} and \ref{sec:2.C} relate to viscoelastic liquids that possess a well defined elasto-capillary regime that can be well-described by the Oldroyd-B fluid. 

However, it is desirable in any rheometric technique to be able to perform measurements without \textit{a priori} selection of a constitutive model. We thus also show how to evaluate bounds on the apparent transient extensional viscosity $\eta_E^+$ as a function of the (time-varying) extensional stress difference $\Delta \sigma(t)$ in the thinning filament, which are agnostic to the material's constitutive model. The key equations required to evaluate these bounds are established in \S \ref{sec:2.D}.

\subsection{Minimum measurable viscosity}\label{sec:2.A}

The minimum filament radius can exhibit a variety of different dynamics depending on the balance between inertial, viscous, capillary and (in the case of viscoelastic fluids) elastic forces. The dimensionless Ohnesorge number $Oh = t_{\nu}/t_R=\eta_0/(\rho \varGamma R_0)^{1/2}$ is used to determine the ratio between viscous and inertial forces when capillarity drives a flow, where $t_{\nu} = \eta_0R_0/\varGamma$ is the viscocapillary time scale and $t_R = (\rho R_0^3/\varGamma)^{1/2}$ is an inertio-capillary time (or Rayleigh time). Here, $\eta_0$ is the zero-shear-rate viscosity, and $\varGamma$ is the liquid-vapour surface tension. For a Newtonian fluid (having a small constant shear viscosity $\eta_0 \equiv \eta$) when $Oh<1$, inertia is the predominant resistance to capillarity-driven thinning dynamics and the minimum radius evolves according to

\begin{equation}
    \frac{R_{\mathrm{ic}}(t)}{R_0} = \alpha \Big( \frac{t_b-t}{t_R} \Big)^{2/3},
    \label{eqn:1}
\end{equation}

\noindent
where $R_{\mathrm{ic}}(t)$ is the predicted functional form for the minimum measured radius $ R_{\mathrm{min}}(t)$ in the inertio-capillary (IC) thinning regime, $\alpha$ is a (configuration-dependent) pre-factor typically between 0.4 and 1 \citep{Day1998,Mckinley2005,Dinic2015}, and $t_b$ is the time to break-up, corresponding to the finite-time singularity when $R_{ic} = 0$ in \eqnref{eqn:1}. Small changes in the value of $\alpha$ affect the precise numerical values of strain-rate in the thinning ligament and the time to break-up, but do not change the 2/3-power scaling with time in \eqnref{eqn:1}.

On the other hand, a Newtonian fluid with large Ohnesorge number $Oh \gtrsim 1$ exhibits visco-capillary (VC) thinning in which the minimum filament radius decreases linearly in time according to

\begin{equation}
    \frac{R_{\mathrm{vc}}(t)}{R_0} = \frac{2X-1}{6} \Big( \frac{t_b-t}{t_{\nu}} \Big).
    \label{eqn:2}
\end{equation}

\noindent
Here, $R_{\mathrm{vc}}(t)$ denotes the predicted functional form for $R_{\mathrm{min}}(t)$ in the VC thinning regime, and $X$ is a constant determined to be identically $X = 0.7127$ for necking of a Newtonian fluid thread in an inertialess Stokes flow of a slender filament at large $Oh$ \citep{Papageorgiou1995,Mckinley2000}. Small changes in the value of $X$ do not change the predicted linear scaling of the radius decay given by \eqnref{eqn:2}.

\citet{Rodd2005} determined a more exact estimate of the Ohnesorge number $Oh$ that delineates the boundaries between IC and VC thinning, by comparing the break-up time $t_b$ in \eqnref{eqn:1} and (\ref{eqn:2}). For IC thinning, the break-up time can be determined by setting $R_{\mathrm{ic}}/R_0 = 1$ and rearranging \eqnref{eqn:1}, demonstrating that $t_b = t_R/\alpha^{3/2}$. Similarly for VC thinning the break-up time is $t_b = 6t_v/(2X-1)$. For VC thinning to be observed, the break-up time from \eqnref{eqn:2} must be greater than that of \eqnref{eqn:1}, which produces the following inequality

\begin{equation}
	Oh = \frac{t_v}{t_R} > \frac{2X-1}{6\alpha^{3/2}}.
	\label{eqn:3}
\end{equation}

\noindent
Given $X = 0.7127$, and typical values of $0.4<\alpha<1$, the transition from IC to VC thinning occurs for Newtonian fluids with viscosities in the range of $Oh=0.07$ to 0.28, depending on the precise value of $\alpha$. \citet{Day1998} predicted analytically that $\alpha = 0.64$ for IC thinning, for which \eqnref{eqn:3} yields the criteria $Oh > 0.14$. However, it should be noted that the exact value of this numerical pre-factor $\alpha$ may depend on several factors, including gravitational effects. For example, the numerical simulations of DoS by \citet{Zinelis2024} demonstrated that the numerical values of $\alpha$ increase systemically with an appropriately defined Bond number $Bo = R_0H \rho g/\varGamma \equiv (H/R_0)\cdot\rho g R_0^2/\Gamma$, measuring the relative importance of capillary and gravitational forces.  

Based on the definition of the Ohnesorge number, $Oh = \eta/(\rho \varGamma R_0)^{1/2}$, the inequality (\ref{eqn:3}) can be rearranged to determine a \textit{dimensional} criteria for the minimum measurable viscosity $\eta$ that can be measured for a Newtonian fluid in a filament thinning experiment,

\begin{equation}
	\eta > \Big( \frac{2X-1}{6\alpha^{3/2}} \Big)\sqrt{\rho \varGamma R_0}.
	\label{eqn:4}
\end{equation}

\noindent
The combination of constants in parenthesis form a single numerical pre-factor $C_{\mathrm{min}}=(2X-1)/(6\alpha^{3/2})$. To illustrate the application of this expression we consider a liquid with a comparable surface tension and density to water ($\varGamma = 70$~mN m$^{-1}$, $\rho = 1,000$ kg m$^{-3}$, $X = 0.7127$, and $\alpha = 0.64$ \citep{Rodd2005}). Substituting these values in \eqnref{eqn:4} the pre-factor is $C_{\mathrm{min}} = 0.14$, and the minimum measurable values of viscosity are $\eta = 22.0$, 29.2 and 37.6 mPa\hspace{0.2em}s for nozzles with radii given by $R_0 = 0.359 \text{ mm}, 0.635 \text{ mm}, \text{and } 1.055 \text{ mm}$, respectively.

\subsection{Minimum measurable relaxation time}\label{sec:2.B}

Non-Newtonian fluids can exhibit various other filament thinning dynamics depending on their rheology. Dilute polymer solution exhibit viscoelastic thinning dynamics depending on the intrinsic Deborah number $De = \tau_E/t_R$, which represents a balance between the viscoelastic and inertio-capillary timescales of the fluid. At a critical time $t^*$ following the start of a capillarity-driven thinning experiment, dilute polymer solutions will undergo a transition from an initial phase of IC or VC thinning (depending on the value of the Ohnesorge number $Oh$) to elastocapillary (EC) thinning, provided there is a sufficient development of polymer stresses within the flow. Dilute solutions of extensible polymers with $De>0$ are expected to exhibit a minimum filament radius that decays exponentially in time according to the following expression, first derived by \citet{Entov1997} for an Oldroyd-B/Hookean dumbbell model;

\begin{equation}
	\frac{R_{\mathrm{ec}}(t)}{R_0} = \Big(\frac{Ec}{2} \Big)^{1/3} \exp\Big(-\frac{(t-t^*)}{3\tau_E}\Big), \hspace{1em} \text{for } t\geq t^*,
	\label{eqn:5}
\end{equation}

\noindent
where $R_{\mathrm{ec}}(t)$ is the predicted functional form of $ R_{\mathrm{min}}(t)$ for elasto-capillary (EC) thinning. Here the elasto-capillary number $Ec = G R_0/\varGamma$ defines the ratio of elastic to capillary stresses, $G = \eta_p/\tau_E$ is the elastic shear modulus of the polymer solution, $\eta_p = \eta_0 - \eta_s$ is the polymeric viscosity, and $\eta_s$ is the solvent viscosity. Note that the elasto-capillary number can also be represented in the form $Ec = (1-\beta)Oh/De$, where $\beta = \eta_s/\eta_0$ is the ratio of the solvent viscosity to zero-shear-rate viscosity. The radius at which the transition to EC thinning occurs, denoted $R_{\mathrm{ec}}(t^*) = R^*$, is

\begin{equation}
    \frac{R^{*}}{R_0} = \Big( \frac{Ec}{2} \Big)^{1/3}.
    \label{eqn:6}
\end{equation}

\noindent
A definition for the transition radius $R^*$ based on the elasto-capillary number $Ec$ was first derived by \citet{Entov1997}, and later refined by \citet{Bazilevskii1997} and \citet{Clasen2006b} to incorporate the additional factor of $2^{1/3}$ in the denominator of \eqnref{eqn:6}. Further refinements to this estimate have been recently discussed by \citet{Gaillard2025}.

Another useful criterion for detecting a transition to EC thinning discussed by \citet{Campo2010}, \citet{Vadillo2012}, and \citet{Sur2018}, is one based on a critical value of the Weissenberg number $Wi$ (a dimensionless representation of the flow strength quantifying the rate at which polymers are being extended within the thinning filament). In filament thinning free-surface flows, the Weissenberg number can be conveniently expressed as

\begin{equation}
	Wi(t) \equiv \tau_E \dot{\varepsilon}(t),
	\label{eqn:7}
\end{equation}

\noindent
where

\begin{equation}
	\dot{\varepsilon}(t) \equiv \frac{-2}{R_{\mathrm{min}}(t)}\frac{dR_{\mathrm{min}}(t)}{dt} = -2\frac{d}{dt}\Big( \ln(R_{\mathrm{min}}) \Big),
	\label{eqn:8}
\end{equation}

\noindent
is the extensional strain rate experienced by a material element near the filament midpoint. For IC thinning, with a minimum radius $R_{\mathrm{ic}}$ described by \eqnref{eqn:1}, the extensional strain rate $\dot{\varepsilon}$ using \eqnref{eqn:8} can be represented as

\begin{equation}
	\dot{\varepsilon}_{\mathrm{ic}}(t) = \frac{4}{3} \frac{1}{(t_b-t)} = \frac{4}{3}\sqrt{\frac{\varGamma \alpha^3}{\rho R_{ic}^3 }},
	\label{eqn:9}
\end{equation}

\noindent
where $t_b$ is a fictitious or extrapolated inertial break-up time from fits of the IC thinning trend at early times -- see \figref{fig:1} for a representation of $t_b$. While for EC thinning, described by the exponential trend of \eqnref{eqn:5}, the extensional strain rate, determined using \eqnref{eqn:8}, is a constant that is independent of time and equal to

\begin{equation}
	\dot{\varepsilon}_{\mathrm{ec}} = \frac{2}{3 \tau_E}.
	\label{eqn:10}
\end{equation}

\noindent
Substituting \eqnref{eqn:10} into \eqnref{eqn:7}, the Weissenberg number in the EC regime is found to be, $Wi_{\mathrm{ec}} = 2/3$. To trigger EC thinning, the rate of deformation in the thinning neck during the early stages of filament thinning must be sufficiently large to extend polymers, in excess of their coil-stretch transition. Assuming the filament thinning dynamics exhibit a transition from IC to EC thinning, the Weissenberg number must satisfy

$$ Wi(t^*) = \frac{4}{3}\frac{\tau_E}{(t_b-t^*)} \geq \frac{2}{3}.$$

\noindent
Rearranging the above inequality to find a constraint on the fluid relaxation time $\tau_E$ yields

$$\tau_E \geq \frac{1}{2}(t_b-t^*) = \frac{t_R}{2}\Big( \frac{R^*}{\alpha R_0} \Big)^{3/2}, $$

\noindent
or equivalently (by dividing through by the Rayleigh time $t_R$), the intrinsic Deborah number of the fluid sample must exceed

$$De \geq \frac{1}{2}\Big( \frac{R^*}{\alpha R_0} \Big)^{3/2}.$$	

\noindent
However, the transition radius $R^*$ also depends on the fluid elasticity \citep{Gaillard2025}, as demonstrated from \eqnref{eqn:6}. If the definition for the transition radius $R^*/R_0$ based on \eqnref{eqn:6} is incorporated in the above inequality, the criterion for the minimum intrinsic Deborah number simplifies to

\begin{equation}
    De \geq \frac{1}{2\alpha} \Big( (1-\beta)Oh \Big)^{1/3}.
    \label{eqn:11}
\end{equation}

\noindent
In dimensional terms this can be represented as

\begin{equation}
    \tau_E \geq \frac{1}{2\alpha} \Big( \frac{\rho \eta_p R_0^4}{\varGamma^2} \Big)^{1/3}.
    \label{eqn:12}
\end{equation}

\noindent
Therefore, \eqnref{eqn:11} and (\ref{eqn:12}) provide theoretical values for the minimum Deborah number and corresponding relaxation time for DoS rheometry. Estimating the numerical values for these limits requires \textit{a priori} measurements of the shear rheology of a complex fluid, for determination of the viscosity ratio $\beta = \eta_s/\eta_0$ and the Ohnesorge number $Oh$ in \eqnref{eqn:11} as well as the polymer contribution to viscosity $\eta_p = (\eta_0 - \eta_s)$ that appears in \eqnref{eqn:12}.

We also note that certain capillarity thinning techniques, namely ROJER, exhibit a slightly different exponential thinning rate in $R_{\mathrm{min}}(t)$, given by \eqnref{eqn:5}. For ROJER, the minimum filament radius decays exponentially with time according to $R_{\mathrm{ec}}(t) \sim \exp(-t/2\tau_E)$ due to additional tension in the filament along the axial direction that is not present in CaBER or DoS \citep{Mathues2018}. This revised scaling results in a slightly larger extensional strain rate in the EC regime $\dot{\varepsilon}_{\mathrm{ec}} = \tau_E^{-1}$ compared to \eqnref{eqn:10}, as well as a characteristic Weissenberg number, $Wi_{\mathrm{ec}}=1$. Therefore, for a transition to EC thinning ROJER must satisfy the criteria $Wi(t^*)\geq1$, effectively resulting in a rescaling of \eqnref{eqn:11} and \eqnref{eqn:12} by a constant factor of $3/2$.   

Model-agnostic expressions for the transient extensional viscosity of an unknown complex fluids are considered next, and this requires a careful considerations of the hardware limits of a typical DoS rheometer configuration.

\subsection{Experimental operating limits}\label{sec:2.C}

\begin{figure}
    \centering
    \includegraphics{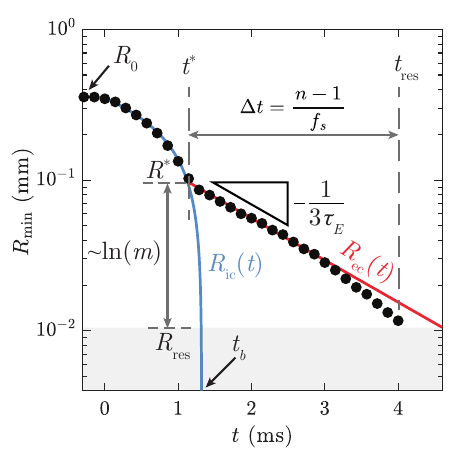}
    \caption{A sample measurement of the minimum filament radius $R_{\mathrm{min}}$ versus time $t$ (represented by the black symbols) for an $M_v = 8.1$ MDa aqueous PEO solution with $c/c^* = 0.06$, for a pendant droplet expelled from a nozzle with $R_0 = 0.359$ mm. The gray shaded region represents the lowest measurable filament radius or $R_{\mathrm{min}}<R_{\mathrm{res}}$. The blue and red lines denote the time-dependent evolution in the minimum filament radius predicted in the inertio-capillary (IC) thinning regime $R_{\mathrm{ic}}$ from \eqnref{eqn:1} and in the elasto-capillary (EC) thinning $R_{\mathrm{ec}}$ domain from \eqnref{eqn:5}, respectively.}
    \label{fig:1}
\end{figure}

In addition to theoretical limitations resulting from the fluid dynamics of filament thinning, accurate measurements of the extensional relaxation time $\tau_E$ can be limited by the DoS experimental hardware \citep{Sur2018}. In the following section a criterion based on the \textit{spatial} and \textit{temporal} resolution of a DoS experimental setup is derived. A sample measurement of the minimum filament radius $R_{\mathrm{min}}(t)$ for a PEO solution dripping from a nozzle with $R_0 = 0.359$ mm radius is shown in \figref{fig:1} and used as a test case for illustrating some of the relevant parameters. The blue solid line in \figref{fig:1} denotes the IC thinning regime $R_{\mathrm{ic}}(t)$, given by \eqnref{eqn:1}, with a fitted prefactor of $\alpha = 0.743$, a Rayleigh time of $t_R = 0.88$ ms, and a break-up time of $t_b = 1.30$ ms. The red solid line represents the EC thinning regime $R_{\mathrm{ec}}(t)$, described by \eqnref{eqn:5}, with an extensional relaxation time determined from an exponential fit to be $\tau_E = 0.52$ ms (or a Deborah number of $De = 0.59$). 

Remarks pertaining to the spatial and temporal resolution of a capillarity-driven extensional rheometer are listed as follows.

\begin{enumerate}
\item \textbf{Spatial resolution:} Firstly, we note that a DoS setup is incapable of measuring the filament radius below an instrument-specific value we define as $R_{\mathrm{res}}$. Values of the filament radius that are below the resolution limit $R_{\mathrm{min}} < R_{\mathrm{res}}$ and cannot be measured, are represented by the grey shaded region in \figref{fig:1}. Generally this resolution limit is prescribed by the camera resolution, i.e., features less than 1 pixel or $\hat{R}_{\mathrm{res}} = 1$ pixel cannot be discerned, regardless of subpixel interpolation algorithms. Throughout this paper, the caret symbol $\hat{\cdots}$ is used to denote the magnitude of spatial variables reported in units of pixels. This limit can be converted to physical units (in this case millimeters) in the object plane using the digital resolution of the imaging system
	
$$R_{\mathrm{res}} = \frac{\hat{R}_{\mathrm{res}}}{\mathcal{D}} = \frac{\hat{R}_{\mathrm{res}} \varDelta_{\mathrm{pix}}}{\mathcal{M}}.$$
	
\noindent
where the digital resolution is $\mathcal{D} = \mathcal{M}/\varDelta_{\mathrm{pix}}$, $\mathcal{M}$ is the imaging magnification, and $\varDelta_{\mathrm{pix}}$ is the physical size of one pixel on the camera sensor (for a typical high speed camera this is around 10 \SI{}{\micro\meter} pixel$^{-1}$ $\leq\varDelta_{\mathrm{pix}}\leq$ 20~\SI{}{\micro\meter}~pixel$^{-1}$). Therefore, increasing the image magnification $\mathcal{M}$ (e.g., by using a higher magnification lens) can reduce $R_{\mathrm{res}}$ and permit measurements of smaller filament radii, $R_{\mathrm{min}}(t)$. The resolution enhancement is, of course, additionally limited by a fixed lower bound corresponding to the diffraction limit of the imaging system (typically $\mathcal{O}$(1 \SI{}{\micro\meter}) for white light illumination).

Assuming the fluid exhibits a transition from IC to EC thinning (like the sample shown in \figref{fig:1}), there is also a \textit{spatial dynamic range}

\begin{equation}
    m = \frac{R^*}{R_{\mathrm{res}}}
    \label{eqn:13}
\end{equation}

\noindent
that defines the range of useful values of the filament radius $R_{\mathrm{min}}(t)$ to which \eqnref{eqn:5} can be fitted and from which an extensional relaxation time can be extracted. Here, $m$ is greater than unity, provided the transition radius is $R^* > R_{\mathrm{res}}$. On a semi-logarithmic representation of $R_{\mathrm{min}}$, the spatial dynamic range can be represented as a distance along the ordinate axis

$$\ln(m) = \ln(R^*)-\ln(R_{\mathrm{res}}),$$

\noindent
as shown in \figref{fig:1}. Therefore, the spatial dynamic range $m$ can be increased by augmenting the spatial resolution or image magnification, i.e., $m \sim R_{\mathrm{res}}^{-1} \sim \mathcal{M}$. For the sample measurement shown in \figref{fig:1}, the spatial dynamic range is $m = 10$.

\item \textbf{Temporal resolution:} To regress experimental data obtained in the EC regime to \eqnref{eqn:5}, a finite number (denoted $n$) of discrete data points are required. Therefore, the duration of the EC thinning regime can also be represented as $\Delta t = (n-1)/f_s$. For the sample measurements shown in \figref{fig:1}, collected at an acquisition rate of $f_s = 4,000$ frames per second (fps), there are $n = 21$ data points in the EC regime, before the spatial resolution limit is encountered.

\end{enumerate}

\noindent
Provided the fluid exhibits an elasto-capillary thinning process with an exponentially decaying radius described by \eqnref{eqn:5}, the ratio between the \textit{spatial resolution} criteria, $\ln(m)$, and the \textit{temporal resolution} criteria, $\Delta t = (n-1)/f_s$ will equal $(3\tau_E)^{-1}$, as shown in \figref{fig:1}. When rearranged, a constraint for the minimum measurable relaxation time corresponding to these experimental criteria is

\begin{equation}
	3\tau_E \geq \frac{n-1}{f_s \ln(m)}.
	\label{eqn:14}
\end{equation}

\noindent
\eqnref{eqn:14} can be rearranged in several different forms to assist experimentalists with estimating the required frame rate $f_s$ of the imaging system, or the minimum values of the measurable relaxation time $\tau_E$. Assuming an experimentalist has some \textit{a priori} estimate of the relaxation time $\tau_E$, the minimum frame rate needed for successful DoS measurements can be established by rearranging \eqnref{eqn:14} such that

\begin{equation}
	f_s \geq \frac{n-1}{3\tau_E\ln(m)}.
	\label{eqn:15}
\end{equation}

\noindent
Provided $\tau_E$ is known, an estimate for $R^*$ and the spatial dynamic range $m$ can be made using \eqnref{eqn:6}. Once the number of data points $n$ to be fitted is selected (ideally $n \gtrsim \mathcal{O}(10)$), \eqnref{eqn:15} can be used to determine the minimum acquisition rate $f_s$ for fitting the data within the EC regime. 

Alternatively, if the camera frame rate $f_s$ is fixed and the relaxation time $\tau_E$ is unknown, \eqnref{eqn:14} can be rearranged to demonstrate that the minimum measurable relaxation time $\tau_E$ for a given experimental configuration is

\begin{equation}
    \tau_E \ln(m) \geq \frac{n-1}{3f_s}.
    \label{eqn:16}
\end{equation}

\noindent
Here, the spatial dynamic range $m$ is kept on the left hand side of the inequality, because it also depends on the level of viscoelasticity in the unknown test sample (note that, $m \sim R^* \sim \tau_E^{-1}$) as per \eqnref{eqn:6} and \eqnref{eqn:13}. Therefore, \eqnref{eqn:16} is an implicit inequality that can be solved to determine a minimum measurable relaxation time $\tau_{E,\textrm{min}}$. For brevity, a simplified explicit form of (\ref{eqn:16}) is not shown here, but provided in Appendix \ref{app:A}.

Note that for ROJER, the ratio between the criteria for spatial and temporal resolution, i.e., $\ln(m)/\Delta t$, is instead equal to $(2 \tau_E)^{-1}$, as shown by \citet{Mathues2018}. Therefore, similar to the analytical criteria of \eqnref{eqn:11} and \eqnref{eqn:12}, the experimental criteria \eqnref{eqn:15} and \eqnref{eqn:16} must be multiplied by a constant factor of $3/2$ when considering jet-based rheometry. This implies that the minimum measurable extensional relaxation time for ROJER is somewhat larger than DoS; however, this factor is relatively small (a 50 \% increase) and ROJER can achieve effectively large sampling rates $f_s$ by using the stroboscopic technique to capture multiple filament thinning events in space (as they advect with the jetting velocity) and time \citep{Keshavarz2015}. This explains why both ROJER and DoS are capable of measuring an extensional relaxation time as small as 10-100 \SI{}{\micro\second} \citep{Keshavarz2015,Keshavarz2016,Sur2018}.

Although \eqnref{eqn:15} and (\ref{eqn:16}) are useful inequalities for establishing the necessary image acquisition rate $f_s$ or minimum measurable relaxation time $\tau_{E,\textrm{min}}$ of a DoS measurement, the user-selected choice of a value of $n$ makes these inequalities subjective. We recommend the number of data points be $n > 10$ for statistically robust fitting of the EC regime. However, to avoid imposing our specific recommendation on the value of $n$, we also propose a ``figure of merit'' that we henceforth refer to as the \textit{filament capture rate} (FCR), which can be reported by DoS practitioners, defined as

\begin{equation}
    \mathrm{FCR} = f_s \ln(m).
    \label{eqn:17}
\end{equation}

\noindent
Here, the FCR (with units of s$^{-1}$) is derived from the observation that $n/\tau_E \sim f_s \ln(m)$ in \eqnref{eqn:14}. This expression reinforces the key idea that a large value of the FCR implies more data points $n$ and better resolution of the EC regime for a fluid with a given relaxation time $\tau_E$. To improve the FCR, an experimentalist could (a) increase the image acquisition rate $f_s$ or (b) increase the image magnification $\mathcal{M}$. From the functional form of \eqnref{eqn:17} it is clear that the FCR is more sensitive to adjustments in temporal resolution ($\mathrm{FCR} \sim f_s$) compared to spatial resolution (since $\mathrm{FCR} \sim \ln \mathcal{M}$). Therefore, a DoS experiment benefits more from a faster camera with a larger frame rate compared to a higher magnification lens.

One limitation of this analysis is it neglects the effects of finite extensibility and assumes that for all times $t^* < t < t_{\mathrm{res}}$ the dynamics faithfully follow the exponential EC thinning process defined by \eqnref{eqn:5}. This assumption is not always valid, even for the present measurements -- for example, \figref{fig:1} demonstrates that measurements of $R_{\mathrm{min}}(t)$ begin to deviate from $R_{\mathrm{ec}}$ at times $t \gtrsim 3$ ms. The effects of finite extensibility can be captured empirically using the functional form proposed by \citet{Anna2001} or using the linear asymptotic result for the FENE-P model established by \citet{Mckinley2005} and implemented by \citet{Oliveira2006} and \citet{Dinic2019}. Other nonlinear regeressions to the measured data can also be preformed using asymptotic analysis of 1-D models such as the Giesekus and FENE-P constitutive equations which give rise to analytical solutions \citep{Torres2014,Wagner2015}. For the sake of simplicity we avoid using these more complex expressions; however, it is worth noting that finite extensibility does effectively reduce the duration $\Delta t$ of the EC regime, implying a larger filament capture rate FCR is needed to capture the complete thinning dynamics.

\subsection{Model agnostic operating limits}\label{sec:2.D}

In \S\S \ref{sec:2.A}, \ref{sec:2.B}, and \ref{sec:2.C}, we derived expressions for the theoretical and experimental minimum viscosities and extensional relaxation times that can be measured using capillarity-driven extensional rheometry. However, deriving these minimum limits required recourse to specific constitutive models, such as a Newtonian fluid model (for which the viscosity can be measured), or the Oldroyd-B/Hookean dumbbell model that is appropriate for dilute polymer solutions, to derive a lower bound on the extensional relaxation time that can be measured. In the following section we provide more generalized limits for DoS rheometry that constrain the range of the transient extensional viscosity $\eta_E^+$ and that are agnostic of the material's constitutive model.

In any capillarity-driven extensional rheometer (CaBER, SRM, DoS, or ROJER), a time-varying tensile stress difference

\begin{equation}
    \Delta \sigma(t) \equiv \sigma_{zz}(t)-\sigma_{rr}(t) = \frac{\varGamma}{R_{\mathrm{min}}(t)},
    \label{eqn:18}
\end{equation}

\noindent
drives the thinning of the fluid ligament (assuming the filament is slender so that the axial curvature $\kappa_z$ of the liquid-vapour interface in the $z-$direction is negligible compared to the  circumferential curvature $\kappa_r = R_{\mathrm{min}}^{-1}$) \citep{Spiegelberg1996,Clasen2006b}. An instantaneous apparent transient extensional viscosity of the material (agnostic of its underlying constitutive equation) can be defined unambiguously as

\begin{equation}
    \eta_E^+(t) \equiv \frac{\Delta \sigma(t)}{\dot{\varepsilon}(t)} = - \frac{1}{2}\frac{\varGamma}{dR_{\mathrm{min}}/dt},
    \label{eqn:19}
\end{equation}

\noindent
where the extensional strain rate $\dot{\varepsilon}(t)$ is defined in \eqnref{eqn:8}. The transient extensional viscosity arises naturally due to the forcing of capillarity which sets the time-evolving magnitude of the tensile stress difference $\Delta \sigma(t)$. We re-emphasize here that the definition in \eqnref{eqn:19} is more completely referred to as an \textit{apparent extensional viscosity} \cite{Mckinley2005}, because neither the applied stress difference, nor the strain rate are generally constant during the filament thinning process. The sole exception to this is, of course, the elastocapillary balance described by \eqnref{eqn:5} and \eqnref{eqn:10} which does establish a constant strain rate in the cylindrical filament corresponding to $Wi = 2/3$. Since capillarity-driven thinning processes are inherently stress-driven, cf. \eqnref{eqn:18}, it is natural to seek bounding expressions for the transient extensional viscosity as a function of the imposed tensile stress difference. These ``operating limits'' are analogous to the operating limits for the calculated shear viscosity $\eta$  as a function of the imposed shear rate $\dot{\gamma}$, derived for steady shear rheometry by \citet{Pipe2008} and \citet{Ewoldt2014}.

For a typical capillarity-driven extensional rheometer, where measurements may be affected by gravity, fluid inertia, amongst other effects, we can identify six distinct limitations on measurements of the transient extensional viscosity $\eta_E^+(t)$ versus the tensile stress difference $\Delta \sigma(t)$. More detailed derivations of these limits are provided in Appendix \ref{app:B}. Two limitations are attributed to other forces that can disrupt the desired balance between the extensional material stress ($\eta_E^+ \dot{\varepsilon}$) generated in the thinning filament and the capillarity-driven tensile stress difference ($\Delta \sigma = \varGamma/R_{\mathrm{min}}$) that was rearranged to produce \eqnref{eqn:19}. The remaining four limits are attributed to experimental constraints. 

We begin by listing the constraints pertaining to the thinning dynamics (see Appendix \ref{app:B} for details of derivations):

\begin{enumerate}
    \item \textbf{Gravitational drainage:} To assume that gravitational drainage has little influence on the thinning dynamics, the stress difference due to capillarity must exceed the gravitational stress acting in the axial direction which will drive sagging in the filament, i.e., we require

    \begin{equation}
        \Delta \sigma(t) \gtrsim \rho g H.
    \label{eqn:20}
    \end{equation}

    \item \textbf{Inertio-capillary thinning:} For the flow in the thinning fluid ligament to be dominated by material stresses (here referring to an extensional material stress difference $\Delta \sigma \equiv \eta_E^+ \dot{\varepsilon}$), and not set by a simple inviscid balance between fluid inertia and capillarity, the extensional viscosity must satisfy
    
    \begin{equation}
        \eta_E^+(t) \gtrsim 2\rho^{1/2}\varGamma  (\Delta \sigma)^{-1/2}.
    \label{eqn:21}
    \end{equation}

    Substituting for the tensile stress difference $\Delta \sigma = \varGamma/R_{\mathrm{min}}(t)$ we can also rearrange this to be,
    $$\eta_E^+(t) \gtrsim 2 \sqrt{\rho \varGamma R_{\mathrm{min}}(t)}.$$
   
\end{enumerate}

\noindent
Experimental resolution constraints on capillarity-driven extensional rheometry can be separated into limitations caused by \textit{spatial} and \textit{temporal} resolution respectively, similar to those discussed in \S \ref{sec:2.C}. Limitations due to spatial resolution include:

\begin{enumerate}
    \setcounter{enumi}{2}
    \item \textbf{Smallest measurable radius:} The smallest radius that can be measured in a DoS experiment is $R_{\mathrm{min}} \geq R_{\mathrm{res}}$. Therefore, a bound on the largest tensile stress difference that can be reliably measured in a capillarity-driven thinning experiment is  
    \begin{equation}
        \Delta \sigma(t) \leq \frac{\varGamma}{R_{\mathrm{res}}}.
    \label{eqn:22}
    \end{equation}

    \item \textbf{Largest useful radius:} Generally a DoS experiment only measures the time-evolving filament radius after a clear ``neck'' has been established from the initial pendant drop. This corresponds to $R_{\mathrm{min}} \leq R_0$. Hence, the smallest measured stress difference that can be reliably inferred is

    \begin{equation}
        \Delta \sigma(t) \geq \frac{\varGamma}{R_0}.
    \label{eqn:23}
    \end{equation}
   
\end{enumerate}

\noindent
The full spatial dynamic range of a DoS experiment ($\Delta \sigma_{\mathrm{max}}/\Delta \sigma_{\mathrm{min}}$) can be determined by taking the ratio of the two previously listed criteria for spatial resolution, i.e., corresponding to the full dynamic range $m_0 = R_0/R_{\mathrm{res}}$. 

Lastly, two constraints due to temporal resolution are provided below:

\begin{enumerate}
    \setcounter{enumi}{4}
    \item \textbf{Largest measurable extensional strain rate:} In Appendix \ref{app:B} it is shown that the largest measurable extensional strain rate depends on the sampling rate and spatial dynamic range according to
    
    $$\dot{\varepsilon}(t) \leq \mathrm{FCR}_0$$
    
    \noindent
    where $\mathrm{FCR}_0 = f_s \ln(m_0)$ is the filament capture rate based on the full spatial dynamic range $m_0 = R_0/R_{\mathrm{res}}$. Therefore, given \eqnref{eqn:19}, this constraint corresponds to a minimum measurable extensional viscosity of

    \begin{equation}
        \eta_E^+(t) \geq \frac{\Delta \sigma}{\mathrm{FCR}_0}.
    \label{eqn:24}
    \end{equation}

    \item \textbf{Minimum resolvable extensional strain rate:} The smallest resolvable extensional strain rate depends on the maximum accessible time of observation,
    
    $$\dot{\varepsilon}(t) \geq \frac{1}{t_{\mathrm{obs}}},$$

    \noindent
    where the observation time $t_{\mathrm{obs}}$ is presumably long in duration, $t_{\mathrm{obs}} \gg f_s^{-1}$. Given \eqnref{eqn:19}, this minimum tolerable extensional strain rate corresponds to a maximum resolvable extensional viscosity of

    \begin{equation}
        \eta_E^+(t) \leq t_{\mathrm{obs}}(\Delta \sigma).
    \label{eqn:25}
    \end{equation}

    \noindent
    The longest observational time $t_{\mathrm{obs}}$ is often set by extrinsic experimental factors such as heat and mass transfer that gives rise to drying, solidification or formation of a ``skin layer'' \citep{Colby2023}. For illustrative purposes, consider filament thinning of a sample with a volatile solvent. The important time scale is the characteristic time scale for solvent evaporation to the ambient surroundings. This may be estimated from measurements of mass transfer,
    
    $$t_{\mathrm{obs}} = t_{\mathrm{evap}} = \frac{R_0}{h_m},$$
    
    \noindent
    where $h_m$ is the mass transfer coefficient of the solvent as it evaporates into the ambient surroundings, with dimensions of $[\mathrm{L}][\mathrm{T}]^{-1}$ \citep{Tripathi2000}. The careful design of environmental control chambers can dramatically reduce the rates of mass transfer (and heat transfer) to the environment and thus extend the bound in \eqnref{eqn:25} set by the observational time scale \citep{Robertson2022,Zhang2022}.
   
\end{enumerate}

\section{Experimental methods}\label{sec:3}

To test some of the derived operating limits, a bespoke DoS setup was developed for experimentally measuring the thinning dynamics of low-viscosity dilute polymer solutions. Details of the experimental DoS configuration are provided in \S \ref{sec:3.A}, while \S \ref{sec:3.B} provides details pertaining to the formulation of the viscoelastic test fluids that will be used to probe the minimum measurable relaxation time $\tau_E$ derived in \S \ref{sec:2.B} and the experimental limitations discussed in \S \ref{sec:2.C}.

\subsection{Dripping-onto-Substrate setup}\label{sec:3.A}

\begin{figure*}
    \centering
    \includegraphics[width=\textwidth]{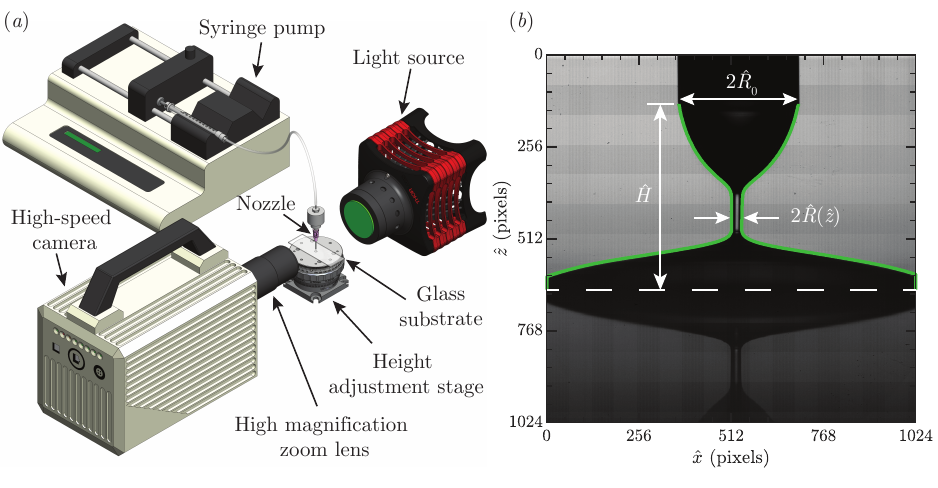}
    \caption{(\textit{a}) Annotated three-dimensional model of the experimental DoS setup, and (\textit{b}) a sample image taken from the high-speed camera for an 8 MDa PEO solution, with $c = 0.14c^*$, $R_0 = 0.635$ mm, and $\varLambda = 1.5$. The green contour line in (\textit{b}) represents the radial profile $\hat{R}(\hat{z})$ extracted from the edge detection algorithm. The white dashed line in (\textit{b}) represents the location of the glass substrate.}
    \label{fig:2}
\end{figure*}

An illustration of the experimental dripping-onto-substrate setup is shown in figure \figref{fig:2}(\textit{a}). A syringe pump (NE-1000, New Era Pump Systems Inc.) is used to dispense a pendant drop of liquid at the tip of a blunt-end nozzle with an outer radius $R_0$. Three different nozzle radii of $R_0 = 0.359 \text{ mm}, 0.635 \text{ mm}, \text{and } 1.055 \text{ mm}$ were considered. Liquid is only contained within the nozzle of the flow setup, while the syringe, and tubing are filled with mostly air (thus requiring only about 100 \SI{}{\micro\liter} of test liquid) to avoid unnecessary straining and potential degradation of the dispensed liquid sample \citep{Joseph2025}. The liquid is dispensed in discrete increments until the bottom of the pendant drop is at a desired height $H$ from the tip of the nozzle. Different heights $H$ were considered for each nozzle radius $R_0$, and the aspect ratio $\varLambda = H/(2R_0)$ for each condition is listed in \tabref{tab:1}. A hydrophilic glass substrate (VWR\textregistered, plain $25 \times 75 \times 1 \text{ mm}^3$ soda lime glass microscope slide) is placed on top of a height adjustable platform with an axial resolution of 10 \SI{}{\micro\meter} (LJ750, Thorlabs Inc.). The position of the substrate is gradually adjusted until it is brought into contact with the bottom of the liquid drop. Upon contact, the droplet wets the substrate, and an unstable liquid bridge configuration is formed. The equilibrium contact angle $\theta$ between the droplet and the glass slide was approximately $\theta = 35^{\circ}$ for all fluids. The components of the experimental setup are adapted from the investigations of \citet{Warwaruk2023,Warwaruk2024} and are similar to those used in \citet{Dinic2015, Dinic2017}. However most of the previously listed investigations \citep{Dinic2015,Dinic2017,Warwaruk2023} used a constant nozzle radius of $R_0 = 1.27$ mm and $\varLambda = 3$, while the present experiments also explore different nozzle radii $R_0$ and aspect ratios $\varLambda$.

Images of the rapidly thinning liquid filament are recorded using a high-speed camera (Fastcam SA-5, Photron) and a collimated high-power (3 W) light-emitting diode (SOLIS-525C, Thorlabs Inc.) that provides back-light illumination. The camera consisted of a $1024 \times 1024$ pixel CMOS (complementary metal-oxide semiconductor) sensor with pixels that were $\varDelta_{\mathrm{pix}} = 20$ \SI{}{\micro\meter} pixel$^{-1}$ in size with a bit-depth of 12-bit. A high-magnification lens (LAOWA, Venus Optics), in combination with a $2\times$ teleconverter, was used to achieve a magnification between $2.8 \leq \mathcal{M} \leq 8.2$, depending on the nozzle radius $R_0$. Here, the magnification $\mathcal{M}$ was adjusted such that the size of the nozzle $\hat{R}_0$ (reported in pixels), and the relative size of the droplet in each image, was approximately the same ($\hat{R}_0 \approx 150$ to 165 pixels) for all cases. Recall from \S \ref{sec:2.C} that the caret symbol $\hat{\cdots}$ is used to denote the magnitude of spatial variables reported in units of pixels. The exact values of magnification are listed in \tabref{tab:1} for each nozzle radius $\hat{R}_0$. An important design parameter, discussed previously in \S \ref{sec:2.C}, is the digital resolution $\mathcal{D} = \mathcal{M}/\varDelta_{\mathrm{pix}}$ that can be used to convert feature sizes in pixels to physical units (in this case millimeters) in the object plane; for example the nozzle radius in millimeters is $R_0 = \hat{R_0}/\mathcal{D}$. Values of the digital resolution $\mathcal{D}$ are provided in \tabref{tab:1}. The image acquisition rate $f_s$ of the camera was adjusted between 250 to 14,000 fps, depending on the viscosity of the fluid being measured.

A sample image of a viscoelastic liquid filament is shown in \figref{fig:2}(\textit{b}). Images are shown with respect to a Cartesian coordinate system (in pixel-space), where $\hat{x}$ and $\hat{z}$ represents the horizontal and vertical dimensions along the image respectively. A simple-to-implement edge detection algorithm is used to establish a radial profile of the filament $\hat{R}(\hat{z})$. Monochromatic images are first binarized using the method of \citet{Otsu1979}. Subsequently, the filament diameter $2\hat{R}(\hat{z},t)$ for each axial position $\hat{z}$ in each frame is then determined by indexing the first and last dark pixels within the binarized image. The green contour in \figref{fig:2}(\textit{b}) shows the resulting radial profile $\hat{R}(\hat{z},t)$ from this simple edge detection algorithm. After establishing the radial profile of the filament, the minimum radius $\hat{R}_{\mathrm{min}}(t)$ along the axial direction $\hat{z}$ was determined for each image in the ensemble. Note, that the minimum radius in pixels $\hat{R}_{\mathrm{min}}(t)$ can be easily converted to physical units $R_{\mathrm{min}}(t)$, according to $R_{\mathrm{min}} = \hat{R}_{\mathrm{min}}/\mathcal{D}$. Measurements of $R_{\mathrm{min}}(t)$ can then be used to extract rheological features of the fluid including the time to breakup $t_b$, and an apparent transient extensional viscosity $\eta_E^+(t)$. Additionally by selecting an appropriate constitutive model, the shear viscosity $\eta$ and the dominant relaxation time $\tau_E$ can be extracted from this time-dependent capillarity-driven flow. At least three repeated measurements were collected for each fluid. The uncertainty in the minimum radius $R_{\mathrm{min}}$ is conservatively estimated to be the range in the thrice-repeated measurements of $R_{\mathrm{min}}$ at each instant of time $t$.

\begin{table}
\caption{\label{tab:1}Geometric and imaging parameters for each DoS configuration.}
\begin{ruledtabular}
\begin{tabular}{ccccc}
$R_0$ (mm) &  $\varLambda$ & $\mathcal{M}$  & $\mathcal{D}$ (pixels mm$^{-1}$) & $\hat{R}_0$ (pixels) \\
\midrule
0.359  & 2.5 & 8.2  & 423  & 152  \\
0.635  & 1.5, 2.0, 2.5 & 5.3  & 260  & 165 \\
1.055  & 1, 1.5, 2.0 & 2.8  & 143  & 151 \\
\end{tabular}
\end{ruledtabular}
\end{table}

\subsection{Test fluids}\label{sec:3.B}

To test these derived limitations on DoS rheometry, experiments in the DoS setup discussed in \S \ref{sec:3.A} were performed with dilute aqueous solutions of PEO and PAM. Two different molecular weights of PEO were considered: 2 MDa (372803, Sigma Aldrich) and 8 MDa (372838, Sigma Aldrich). A PAM polymer, with an advertised molecular weight in the range of 17-23 MDa (Flopam AN 934 VHM, SNF Floerger), was also examined. Previous DoS measurements of PEO are in abundance, and therefore, PEO serves as a nice test polymer for comparison with previous investigations \citep{Dinic2015,Sur2018,Robertson2022}. DoS measurements of ultra high molecular weight PAM polymers are less frequently considered -- with the exception of the recent works by \citet{Gaillard2024,Gaillard2025} and \citet{Joseph2025} -- but PAM solutions are often employed in many other investigations of complex fluids, particularly those involving elastically-driven instabilities and turbulent polymer drag reduction \citep{Howe2015,Lacassagne2021,Browne2024,Warwaruk2024,Chen2025}. Therefore, the relaxation time $\tau_E$ is also measured for this particular commercial PAM product with an exceptionally large molecular weight.

\begin{figure}
    \centering
    \includegraphics{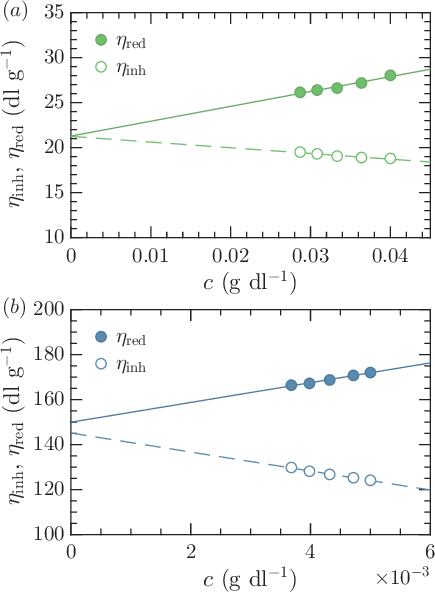}
    \caption{Measurements of the reduced and inherent viscosity for (\textit{a}) PEO $M_v = 8.1$ MDa, and (\textit{b}) PAM with a reported molecular weight in the range of $M_v = 17-23$ MDa. Solid lines represent fits of the reduced viscosity using the Huggins \eqnref{eqn:26a}, while dashed lines are fits of the inherent viscosity using the Kraemer \eqnref{eqn:26b}.}
    \label{fig:3}
\end{figure}

Concentrated polymer solutions (1000 ppm or 0.1 g dl$^{-1}$) were prepared by gently adding solid polymer powder to 2 l of deionized water. The concentrated solutions were then slowly mixed using a roll-mixer for approximately 24 hours, after which the solutions were diluted to the desired concentrations $5 \text{ ppm} \leq c \leq 200 \text{ ppm}$. An Ubbel\"{o}hde capillary viscometer (size 0B, Cannon Instrument) was used to measure the zero-shear-rate viscosity $\eta_0$ of the polymer solutions, using the procedure described in ASTM D445 \citep{ASTMD445}. The reduced viscosity $\eta_{\mathrm{red}} = (\eta_0-\eta_s)/\eta_s c$ and inherent viscosity $\eta_{\mathrm{inh}} = \ln(\eta_0/\eta_s)$, were determined for each fluid, and representative data are shown in \figref{fig:3}. Here $\eta_s$ is the viscosity of the solvent (in this case, deionized water with $\eta_s = 1 \times 10^{-3}$ Pa s), and all fluids were evaluated at a temperature of approximately $T = 21.0^{\circ}\text{C} \pm 0.5^{\circ}\text{C}$. \figref{fig:3}(\textit{a}) and (\textit{b}) show sample measurements of $\eta_{\mathrm{red}}$ and $\eta_{\mathrm{inh}}$ for the higher molecular weight PEO polymer, and PAM, respectively. Values of $\eta_{\mathrm{red}}$ and $\eta_{\mathrm{inh}}$ with respect to $c$, were fit using the relationships of \citet{Huggins1942} and \citet{Kraemer1938},

\begin{subequations}
\begin{align}
	\eta_{\mathrm{red}} & = \frac{\eta_0-\eta_s}{\eta_s c} =  [\eta]+k_H[\eta]^2c + \mathcal{O}(c^2), 	\label{eqn:26a} \\ 
	\text{and} \nonumber \\
	\eta_{\mathrm{inh}} & = \frac{1}{c}\ln\Big(\frac{\eta_0}{\eta_s}\Big) = [\eta]-k_K[\eta]^2c + \mathcal{O}(c^2),
	\label{eqn:26b}
\end{align}
\end{subequations}

\noindent
where the higher-order polynomial terms in the concentration $c$, i.e., $\mathcal{O}(c^2)$, are assumed to be negligible. Here, $[\eta]$ is the intrinsic viscosity, which can be determined based on the double intercept between \eqnref{eqn:26a} and (\ref{eqn:26b}) with the ordinate axis (where the polymer concentration $c \rightarrow 0$ g dl$^{-1}$). The Huggins constant $k_H$ and Kraemer constant $k_K$ represent the slopes of \eqnref{eqn:26a} and (\ref{eqn:26b}), respectively. Fits using \eqnref{eqn:26a} and (\ref{eqn:26b}) are represented with solid and dashed lines, respectively, in \figref{fig:3}. For all fluids, both the Huggins \eqnref{eqn:26a} and Kraemer \eqnref{eqn:26b} show a single self-consistent intercept with the ordinate axis, demonstrating good confidence in the derived value of the intrinsic viscosity $[\eta]$. The intrinsic viscosity $[\eta]$ from combined Huggins-Kraemer fits are listed in \tabref{tab:2} for each polymer.

\begin{table*}
\caption{\label{tab:2}Chemical and rheological features of the test fluids.}
\begin{ruledtabular}
\begin{tabular}{lcccccc}
Polymer  & $[\eta]$ (dl g$^{-1}$) & $M_v$ (MDa) & $c^*$ (g dl$^{-1}$) & $\tau_Z$ (ms) & $c/c^*$ & $\beta$ \\
\midrule
\multirow{ 2}{*}{PEO} & 9.80 & 2.5 & 0.0785 & 0.48 & 0.06, 0.13, 0.25 & 0.95, 0.91, 0.84\\
& 21.3  & 8.1  & 0.0362 & 3.34 & 0.06, 0.14 & 0.95, 0.90 \\
PAM & 148 & 17-23\footnote{Values reported by the polymer manufacturer.} & 0.0052 & - & 0.10, 0.19, 0.38 & 0.93, 0.87, 0.77 \\
\end{tabular}
\end{ruledtabular}
\end{table*}

Comparing the intrinsic viscosity determined for the different polymers listed in \tabref{tab:2}, it is expected that the values of $[\eta]$ generally increases with molecular weight. The intrinsic viscosity $[\eta]$ is expected to scale with molecular weight based on the Mark-Houwink-Sakurada (MHS) expression, $[\eta] = KM_v^a$, where $M_v$ is the viscosity-average molecular weight, and $K$, $a$ are the MHS coefficient and exponent, respectively \citep{Rubinstein2003}. Furthermore, for dilute solutions of flexible polymer chains the MHS exponent can be related to Flory's solvent quality $\nu$, through the expression $a = 3\nu-1$. For PEO, values of the MHS constants have previously been reported as $K  = 50.0\times10^{-5}$ dl mol$^{a}$ g$^{-(a+1)}$ and $a = 0.67$ ($\nu = 0.56$) \citep{Gregory1986,Brandrup1999}. These values of the MHS coefficients are also approximately equal to those used in other investigations of PEO from the same polymer supplier \citep{Tirtaatmadja2006}. The viscosity-averaged molecular weight $M_v$ can be determined by rearranging the MHS equation to give $M_v = ([\eta]/K)^{1/a}$, with the resulting values listed in \tabref{tab:2}. The computed values of the molecular weight $M_v$, 2.5 and 8.1 MDa, are not markedly different from the nominal values of 2 and 8 MDa advertised by the polymer supplier.

For PAM, there are various mixed reportings of the MHS coefficients. For example, \citet{Tam1991} compiled values of the MHS constants across different investigations, demonstrating a range of values between $1.3\times10^{-5}$ dl mol$^{a}$ g$^{-(a+1)}$ $<K<$  $17.0\times10^{-5}$ dl mol$^{a}$ g$^{-(a+1)}$, and $0.66 < a < 0.83$ -- see also \citet{Misra1979,Shawki1979,Brandrup1999}. Considering these different MHS parameters yields a wide range of possible molecular weights for the present PAM system, $M_v \sim \mathcal{O}(10^1)-\mathcal{O}(10^3)$ MDa -- the upper extent being highly unlikely. Hence there would be significant uncertainty in any estimate of the molecular weight for PAM, from our measured value of the intrinsic viscosity. Furthermore, PAM polymers are also known to exhibit a strong propensity for hydrogen bonding that encourage associations with neighboring polymer molecules \citep{Ait1987} -- an effect that is exacerbated for these ultra high molecular weight systems. Many prior investigations have used salt to mitigate the effects of these associations and ensure the PAM molecules adopt a random coiled configuration at equilibrium \citep{Ait1987,Zhang2021,Gaillard2024,Gaillard2025}. However, in the current investigation, the solvent is deionized water, with no additional salts or ions. Overall, given the mixed reporting of the MHS constants and risk of polymer associations, we avoid computing a single value for the molecular weight $M_v$ for PAM. We thus report in \tabref{tab:2} the manufacturer's reported range of molecular weights. Despite these reservations, we note that the measured intrinsic viscosity value $[\eta]$ is still valid for this system, and provides a good indication that in dilute solutions, PAM molecules pervade more volume than the flexible and narrowly-distributed PEO chains. 

Given the measured values of intrinsic viscosity $[\eta]$ for each system, the coil overlap concentration $c^* = 0.77/[\eta]$ was determined according to the expression of \citet{Graessley1980}, and is also listed in \tabref{tab:2} for each polymer. The longest relaxation time for a single infinitely dilute polymer chain in a good solvent can also be calculated using Zimm theory and the following expression 

\begin{equation}
	\tau_Z \simeq \frac{1}{\zeta(3\nu)} \frac{[\eta]M_v\eta_s}{N_Ak_BT} = \frac{1}{\zeta(3\nu)}\frac{K M_v^{a+1}\eta_s}{N_Ak_BT},
	\label{eqn:27}
\end{equation}

\noindent
where $\zeta(3\nu) = \sum_{i=1}^{\infty} 1/(i^{3\nu})$. Values of the Zimm relaxation time $\tau_Z$ are listed in \tabref{tab:2} for the two PEO samples of different molecular weight $M_v$ based on \eqnref{eqn:27}. Given that the reported estimates of the MHS constants, $K$ and $a$, have large variability and that the manufacturer reported molecular weight $M_v$ is highly polydisperse for the PAM polymer, there would be very large uncertainty in any estimates of a ``Zimm time'' for this polydisperse solution, and we thus avoid reporting any value in \tabref{tab:2} beyond our experimentally validated intrinsic viscosity $[\eta]$, and the corresponding calculated value of $c^*$.

Several dilute aqueous solutions (with $c<c^*$) of each polymer were prepared for DoS measurements, the specific values of which are listed in \tabref{tab:2}.  The zero-shear-rate viscosity $\eta_0$ can be determined for each solution using the Huggins \eqnref{eqn:26a} or Kraemer \eqnref{eqn:26b} equations and the intrinsic viscosity $[\eta]$, listed for each polymer in \tabref{tab:2}. For comparison with predictions of dumbbell models an important dimensionless representation for \eqnref{eqn:26a} is $\beta \approx (1+[\eta]c)^{-1}$, where $\beta = \eta_s/\eta_0$ is the dimensionless solvent contribution to the total viscosity ($0<\beta<1$). Values of the dimensionless viscosity $\beta$ are listed in \tabref{tab:2} for each polymer concentration $c/c^*$. 

The surface tension $\varGamma$ of each fluid was measured using a force tensiometer (DCAT-11 Data Physics) equipped with a platinum Du No{\"u}y ring. It is well known that aqueous PEO solutions tend to be weakly surface active, with a lower surface tension $\varGamma$ compared to water \citep{Cooper2002,Rodd2005,Tirtaatmadja2006}. All PEO solutions (regardless of different molecular weight $M_v$ and concentration $c$) were measured to have similar values of surface tension, $\varGamma = 63.0 \pm 0.5$ mN m$^{-1}$. On the other hand, PAM solutions exhibited a surface tension unchanged from the value measured for water, $\varGamma = 72.2 \pm 0.2$ mN m$^{-1}$. The uncertainties represent the standard deviation in the measured values of surface tension for different repeated trials, and concentration $c$ (as well as molecular weight $M_v$ in the case of PEO).

\section{Results}

\subsection{Testing the limits}\label{sec:4.A}

In \S \ref{sec:2.B} a theoretical minimum Deborah number and relaxation time for DoS measurements, given by the inequalities (\ref{eqn:11}) and (\ref{eqn:12}), was established. Furthermore, a minimum measurable relaxation time based on experimental parameters was established in \S \ref{sec:2.C}, and described by the inequality (\ref{eqn:16}). The following results test these limits experimentally by measuring the relaxation time of the dilute PEO and PAM solutions detailed in \S \ref{sec:3.B}. We begin by measuring the relaxation time of the lower molecular weight ($M_v = 2.5$ MDa) PEO solutions in the $R_0 = 0.359$ mm nozzle at different $c/c^*$. However, before discussing the experimental results for these fluids, the theoretical minimum measurable relaxation time, based on \eqnref{eqn:11} and (\ref{eqn:12}), are first calculated.

For the lowest concentration PEO solution with $c/c^* = 0.06$, the viscosity ratio is $\beta = 0.95$, as listed in \tabref{tab:2}, and the Ohnesorge number is $Oh = 7.2 \times 10^{-3}$, as listed in \tabref{tab:3}. Based on fits of the inertio-capillary (IC) regime, the pre-factor was $\alpha = 0.74$ for all of the tested PEO fluids using the $R_0 = 0.359$ mm nozzle (values of which are again listed in \tabref{tab:3}). Therefore, using the inequality (\ref{eqn:11}), the minimum measurable Deborah number is $De \geq 0.05$ for the PEO solutions. Given a Rayleigh time of $t_R = 0.88$ ms, the minimum measurable relaxation time is thus estimated from \eqnref{eqn:12} to be $\tau_E \geq 0.04$ ms. However, as we discussed in \S \ref{sec:2.C}, in addition to this theoretical minimum there are also experimental constraints that limit our ability to measure very small relaxation times. 

\begin{figure*}
    \centering
    \includegraphics[width=\textwidth]{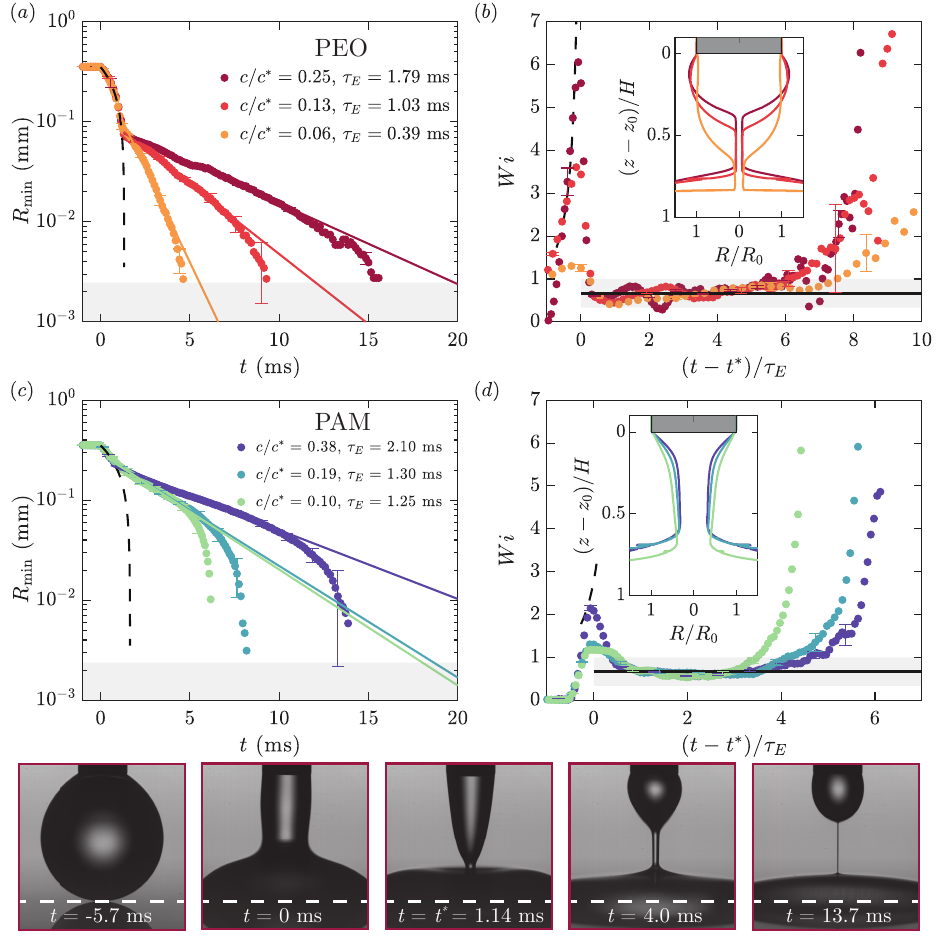}
    \caption{Plots of (\textit{a}) the minimum filament radius $R_{\mathrm{min}}$ versus time $t$ and (\textit{b}) the time-varying Weissenberg number $Wi$ (Eq. \ref{eqn:9}) versus time $(t-t^*)/\tau_E$ for the $M_v = 2.5$ MDa PEO solution at different concentrations $c/c^*$ and in the nozzle radius $R_0 = 0.359$ mm. Similarly (\textit{c}) shows the evolution in the minimum radius $R_{\mathrm{min}}$ versus $t$ and (\textit{d}) the evolution in the Weissenberg number $Wi$ versus $(t-t^*)/\tau_E$ for the $M_v = 17-23$ MDa PAM solution at different $c/c^*$ and $R_0 = 0.359$ mm. Solid coloured lines in (\textit{a}) and (\textit{c}) correspond to fits of the elasto-capillary (EC) regime using \eqnref{eqn:5}, while the black dashed line is a fit of the inertio-capillary (IC) regime for the fluid with the lowest $c/c^*$. More information about the flows is listed in \tabref{tab:3}. The grey shaded region in (\textit{a}) and (\textit{c}) denote $R_{\mathrm{res}} = 0.002$ mm. The black solid line in (\textit{b, d}) represents $Wi = 2/3$ for the EC regime, where the grey region represents a 50 \% confidence region. The dashed black line (\textit{b, d}) represents the analytical expression for Weissenberg number in the IC regime $Wi(t) = \dot{\varepsilon}_{\mathrm{ic}}\tau_E$. Insets in (\textit{b, d}) shows the shape of the filament, coloured according to the different $c/c^*$, while the dark shaded region  represents the nozzle. All profiles in the inset of (\textit{b}) correspond to the same time instance of $(t-t^*)/\tau_E = 3.5$, while the inset in (\textit{d}) corresponds to $(t-t^*)/\tau_E = 2.0$. The \textbf{bottom panels} show images of the droplet thinning for the $c/c^* = 0.25$ PEO solution at different time instances $t$ are shown in the bottom panels. The time $t = 0$ corresponds to the time instance when the minimum radius equals the nozzle radius, $R_{\mathrm{min}} = R_0$.}
    \label{fig:4}
\end{figure*}

\begin{table*}
\caption{\label{tab:3}Dimensional and dimensionless rheological properties for different viscoelastic fluids in DoS experiments using the $R_0 = 0.359$ mm nozzle shown in \figref{fig:4}.}
\begin{ruledtabular}
\begin{tabular}{lccccccc}
Fluid  & $c/c^*$ & $t_R$ (ms) & $Oh$ $(\times 10^{-3})$ & $\alpha$ & $f_s$ (fps) & $\tau_E$ (ms) & $De$\\
\midrule
\multirow{ 3}{*}{PEO, $M_v = 2.5$ MDa} & 0.06 & \multirow{ 3}{*}{0.88} &       7.2 & 0.74 & 9,300 & 0.39 & 0.44 \\
 & 0.14 &  & 7.5 & 0.73 & 7,000 & 1.03 & 1.12\\
 & 0.25 &  & 8.2 & 0.72 & 7,000 & 1.79  & 2.04\\
		
\midrule
		
\multirow{ 3}{*}{PAM, $M_v = 17-23$ MDa} & 0.10  &\multirow{ 3}{*}{0.80} & 6.7 & 0.64 & 12,500 & 1.25 & 1.56 \\
  & 0.14 &  & 7.2 & 0.62 & 9,300 & 1.30 & 1.62\\
  & 0.25 &  & 8.2 & 0.60 & 7,000 & 2.10 & 2.62\\
\end{tabular}
\end{ruledtabular}
\end{table*}

The experimental limitations for DoS rheometry require considerations of both the spatial and temporal resolution of the DoS setup. Regarding \textit{spatial resolution}, we require an estimate for the spatial dynamic range $m = R^*/R_{\mathrm{res}}$. Recall that the digital resolution is $\mathcal{D} = 423$ pixel mm$^{-1}$ for the $R_0 = 0.359$ mm nozzle, as listed in \tabref{tab:1}. Therefore a radial resolution of $\hat{R}_{\mathrm{res}} = 1$ pixel is equivalent to $R_{\mathrm{res}} = \hat{R}_{\mathrm{res}}/\mathcal{D} = 0.002$ mm. To determine an \textit{a priori} estimate of $R^*$, \eqnref{eqn:6} can be utilized. For the lowest concentration PEO solution, the viscosity ratio is $\beta = 0.95$, the Ohnesorge number is $Oh = 7.2 \times 10^{-3}$ and the theoretical minimum Deborah number previously determined was $De = 0.05$. Therefore, an estimate for the elasto-capillary number $Ec = \eta_pR_0/\tau_ER_0 = (1-\beta)Oh/De$ for this low concentration PEO solution is $Ec = 7.2 \times 10^{-3}$, and the transition radius, according to \eqnref{eqn:6}, is $R^* = 0.15R_0 = 0.05$ mm. Using \eqnref{eqn:13}, therefore, the spatial dynamic range is $m = 25$. 

Regarding \textit{temporal resolution}, to be conservative we stipulate $n = 20$ data points are sufficient for measuring and fitting the EC regime. Therefore, using the theoretical minimum relaxation time $\tau_E = 0.04$ ms and $m = 25$ as an input to the inequality (\ref{eqn:15}), an estimate for the minimum necessary frame rate is $f_s \geq 49,000$ fps. Although this large frame rate is achievable (e.g., \citet{Sur2018} used $f_s = 50,000$ fps), a more commonly accessible camera frame rate is $f_s = 10,000$~fps. Using the value $f_s = 10,000$~fps in \eqnref{eqn:16} we can calculate an experimentally constrained minimum measurable relaxation time, which is $\tau_E \geq 0.23$ ms or $De \geq 0.26$ (determined using the explicit expression provided in Appendix \ref{app:A}). Conveniently, this experimental minimum relaxation time is similar in magnitude compared to the \textit{a priori} estimate of the Zimm relaxation time $\tau_Z = 0.48$~ms estimated for the $M_v = 2.5$~MDa PEO solution, provided in \tabref{tab:1}. Therefore, low relaxation times close to the Zimm time $\tau_Z$ are, in principle, measurable for sufficiently dilute solutions of this flexible high molecular weight polymer. 

As previously noted, a metric that encapsulates both the spatial and temporal resolution needed for these DoS measurements is the \textit{filament capture rate}. Given the assumed frame rate of $f_s = 10,000$~fps and a spatial dynamic range of $m = 25$, the filament capture rate is $\mathrm{FCR} = f_s \ln(m) = 32,200$~s$^{-1}$. In summary, we can make two independent estimates of the minimum relaxation time: a theoretical minimum $\tau \geq 0.04$~ms from \eqnref{eqn:12}, and an experimental minimum $\tau_E \geq 0.23$~ms from \eqnref{eqn:16}. 

Profiles of the minimum filament radius $R_{\mathrm{min}}(t)$ are shown in \figref{fig:4}(\textit{a}) for the $M_v = 2.5$~MDa PEO solution. All concentrations ($c/c^* = 0.06, 0.13, 0.25$) demonstrate consistency with the IC trend of \eqnref{eqn:1} where the pre-factor is $\alpha = 0.74$, and the break-up time (for a low viscosity Newtonian fluid) is $t_b = 1.3$~ms. At time $t>t^*$ the PEO solutions exhibit EC thinning dynamics consistent with \eqnref{eqn:5}, and the values of the relaxation time $\tau_E$ reduce as concentration $c/c^*$ decreases. At the lowest concentration of $c/c^* = 0.06$, the relaxation time is calculated from the measurements to be $\tau_E = 0.39$~ms, which is comparable to the theoretical Zimm time $\tau_Z = 0.48$~ms. \figref{fig:4}(\textit{b}) demonstrates the temporal variation in the Weissenberg number $Wi(t)$, established using \eqnref{eqn:6}, for the $2.5$ MDa PEO solutions at different concentrations $c/c^*$. The black solid line and gray shaded region represents $Wi = 2/3 \pm 1/3$, corresponding to the theoretical Weissenberg number for the EC regime, with a 50 \% confidence interval. Generally all PEO solutions shown in \figref{fig:4}(\textit{b}) fall within this conservative confidence interval of Weissenberg number for a duration of about $5\tau_E$ to $7\tau_E$, consistent with the observations of \citet{Dinic2019} who reported that the EC regime was approximately $6\tau_E$ in duration for semi-dilute solutions of PEO. Inertial oscillations are responsible for the deviations between the measured values of Weissenberg number $Wi$ and the theoretical value $Wi = 2/3$ expected for the EC regime, by virtue of the very low Ohnesorge and Deborah numbers of these tests, (cf. \tabref{tab:3}). That being said, the inset axis in \figref{fig:4}(\textit{b}) demonstrates that the shape of the filament in the EC regime (at time $(t-t^*)/\tau_E = 3.5$) is cylindrical for all concentrations $c/c^*$, a necessary requirement for derivation of the 1-D elasto-capillary thinning process described by \eqnref{eqn:5}.

The results of \figref{fig:4}(\textit{a, b}) demonstrate that small values of the relaxation time $\tau_E$, that are on the order of 0.1~ms, and close to the Zimm relaxation time $\tau_Z$ (for a $M_v = 2.5$~MDa PEO solution) can be reliably measured with DoS rheometry. The lowest experimentally measured relaxation time of $\tau_E = 0.39$~ms is also in close agreement and still exceeding the minimum limits predicted from theory ($\tau_E \geq 0.04$~ms using \eqnref{eqn:12}) or experiments ($\tau_E \geq 0.23$~ms using \eqnref{eqn:16}).

In addition to PEO, we also explore the lowest measurable relaxation time $\tau_E$ for an ultra high molecular weight polydisperse drag-reducing polymer, in this case the $M_v = 17-23$~MDa PAM polymer. This fluid has similar values of $\beta$, $Oh$ and $\alpha$ as the PEO polymer, as listed in \tabref{tab:2} and \tabref{tab:3}, and is subject to a similar limitation on relaxation time, i.e., the theoretical and experimental limits of $\tau_E \geq 0.04$~ms and $\tau_E \geq 0.23$~ms. \figref{fig:4}(\textit{c}) demonstrates the minimum filament radius $R_{\mathrm{min}}(t)$ for the PAM solutions at different concentrations $c/c^*$. Regardless of concentration, the PAM solutions exhibit similar IC thinning with a slightly smaller pre-factor of $\alpha = 0.62$ and break-up time of $t_b = 1.6$~ms, as shown by the black dashed line in \figref{fig:4}(\textit{c}). A noticeable difference between the PEO and PAM solutions shown in \figref{fig:4}(\textit{a}) and (\textit{c}) are the critical conditions $t^*$ and $R^*$ corresponding to the transition from IC to EC thinning. All of the PAM solutions exhibit a transition radius of $R^* = 0.28$~mm, that is approximately three times larger than the value of $R^* = 0.09$~mm observed for PEO. This is consistent with the expectation that even a small fraction of very high molecular weight chains can contribute substantially to the elastic stress that develops in the filament during the rapid stretching that occurs in the IC regime \citep{Calabrese2025}. Despite this early transition to EC thinning, the PAM solutions show good agreement with the EC trend of \eqnref{eqn:5}. The relaxation time $\tau_E$ decreases from 2.10~ms to 1.30~ms as $c/c^*$ is lowered from 0.38 to 0.19. However, as $c/c^*$ is lowered further to 0.10, the relaxation time appears to plateau, reflecting a value $\tau_E = 1.25$~ms similar to $\tau_E = 1.30$~ms for $c/c^* = 0.19$. A plateau of relaxation time in the limit of low concentration is consistent with dilute solution theory, implying the effective Zimm time is $\tau_Z \approx 1.25$~ms for this polydisperse PAM polymer. However, we avoid drawing this conclusion without a good comparison with a Zimm time determined from polymer chemistry or linear viscoelasticity. This highlights the need for better characterization of the MHS parameters for these ultra high molecular weight polydisperse PAM systems.

\figref{fig:4}(\textit{d}) demonstrates the evolution in the Weissenberg number $Wi(t)$ with respect to a dimensionless time $(t-t^*)/\tau_E$ for the PAM solutions of different concentration $c/c^*$. The inset axis in \figref{fig:4}(\textit{d}) shows the shape of the filament at a time $(t-t^*)/\tau_E = 2$, corresponding to the middle of the EC regime. Although the low concentration PAM solutions exhibit a region of exponential thinning consistent with the EC regime balance, $Wi = 2/3$, the shape of their filament does not closely resemble a 1-D cylindrical profile (in which the radius $R$ is homogeneous along the axial direction $z$). Interestingly, as the concentration increases from $c/c^* =0.10$ to 0.38, the filament visually appears to become increasingly cylindrical at the same time instance. The durations of the EC regime for the PAM solutions are approximately $3\tau_E$ to $5\tau_E$, which is generally shorter than that observed for PEO solutions shown in \figref{fig:4}(\textit{b}) and reported by \citet{Dinic2019}. \citet{Zinelis2024} showed that viscoelastic filaments with low values of $L^2$ (the maximum squared end-to-end length of a FENE dumbbell) will decay more rapidly compared to those with Hookean springs ($L^2 \rightarrow \infty$). Given the short durations of the EC regime, observed in \figref{fig:4}(\textit{d}), we suspect these PAM solutions are characterized by a lower value of $L^2$, exhibiting a less extensible confirmation within the flow. Clearly these ultra high molecular weight and polydisperse PAM solutions behave quite distinctly to the PEO solutions. To our knowledge, the only other investigation that has used similar ultra-high molecular weight PAM systems for DoS is \citet{Joseph2025}, albeit at a larger concentration ($c/c^* = 0.91$, where $c^* = 0.075$ g dl$^{-1}$) and in a more viscous solvent ($\eta_s = 0.26$ Pa s).

\subsection{Influence of gravitational body force}\label{sec:4.B}

\begin{figure*}
    \centering
    \includegraphics[width=\textwidth]{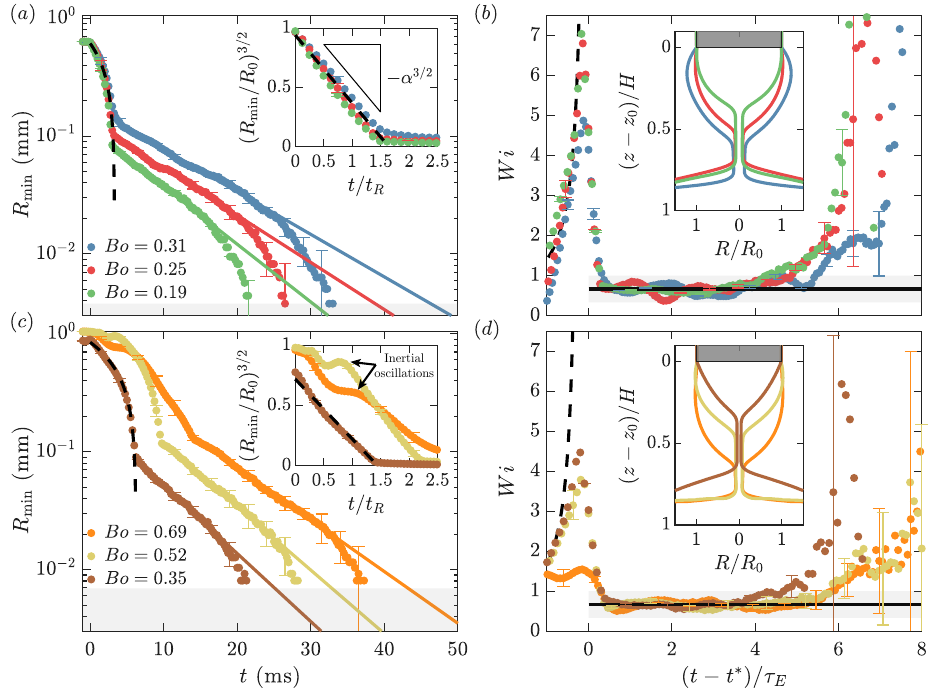}
    \caption{Filament thinning DoS measurements all for a $M_v = 8.1$ MDa, $c/c^* = 0.14$ PEO solution, showing (\textit{a}) the minimum filament radius $R_{\mathrm{min}}$ as a function of time $t$ and (\textit{b}) the time-varying Weissenberg number $Wi(t)$ (Eq. \ref{eqn:9}) as a function of $(t-t^*)/\tau_E$ in the $R_0 = 0.635$ mm nozzle at different aspect ratios $\varLambda$ or Bond numbers $Bo$. Similarly, (\textit{c}) shows $R_{\mathrm{min}}(t)$ and (\textit{d}) $Wi$ versus $(t-t^*)\tau_E$ for the same PEO solution as (\textit{a,b}), but in a $R_0 = 1.055$ mm nozzle, at different $\varLambda$ and $Bo$. More information about each experimental configuration is listed in \tabref{tab:4}. The grey shaded region in (\textit{a}) is bounded by $R_{\mathrm{res}} = 0.003$ mm, and in (\textit{c}) is $R_{\mathrm{res}} = 0.007$ mm. Coloured solid lines in (\textit{a}) and (\textit{c}) correspond to fits of the elasto-capillary (EC) regime using \eqnref{eqn:5}. The dashed black line in represents a fit of the inertio-capillary (IC) regime using \eqnref{eqn:1} for $Bo = 0.25$ for (\textit{a}) and $Bo = 0.35$ in (\textit{c}). The inset axis in (\textit{a, c}) shows the minimum radius rescaled in such a way that the IC prediction given by \eqnref{eqn:1} is linear with a slope of $-\alpha^{3/2}$. The black solid line in (\textit{b, d}) represents thew value $Wi = 2/3$ for the EC regime, where the grey shaded bar represents a 50 \% confidence region. The dashed black line in (\textit{b, d}) represents the analytical expression for the Weissenberg number in the IC regime $Wi(t) = \dot{\varepsilon}_{\mathrm{ic}}\tau_E$. Insets in (\textit{b, d}) shows the different shapes of the filament, coloured according to the different $Bo$, while the dark shaded region denotes the nozzle. All profiles in the inset of (\textit{b}) and (\textit{d}) correspond to the same time instance of $(t-t^*)/\tau_E = 3$.}
    \label{fig:5}
\end{figure*}

The influence of gravity on DoS rheometery is often overlooked. The ratio of gravitational to capillary forces is defined by a dimensionless Bond number, represented as $Bo = R_0H/\ell_c^2$, where $\ell_c = (\varGamma/\rho g)^{1/2}$ is the capillary length (a representative length scale balancing surface tension to gravity). In DoS rheometry, it is typically assumed that the measured values of relaxation time $\tau_E$ do not depend on the nozzle height $H$ or radius $R_0$, and thus by extension on the aspect ratio $\varLambda$ and Bond number $Bo$ \citep{Dinic2015}. Yet, surprisingly few investigations have explored the effect of varying the Bond number $Bo$ on the filament thinning dynamics. One exception is the time-dependent numerical simulations of DoS presented by \citet{Zinelis2024} that used the Oldroyd-B and FENE-P constitutive equations, to directly explore the influence of Bond number $Bo$ on the thinning dynamics. Their simulations over the range $0 \leq Bo \leq 1$ found that the Bond number had no effect on the rate of filament thinning within the EC regime (i.e., on determination of the extensional relaxation time $\tau_E$), but greatly modified the IC regime, in particular the value of the pre-factor $\alpha$ and the condition characterizing the transition from IC to EC thinning, $t^*$ and $R^*$. \citet{Zinelis2024} observed a scaling between the IC pre-factor and the Bond number equivalent to, $\alpha \sim Bo^{2/3}$. Furthermore, \citet{Zinelis2024} showed that the transition radius $R^*$ increases and the transition time $t^*$ decreases as the Bond number was increased.

\begin{table*}
\caption{\label{tab:4} Measurements for the same $M_v = 8.1$ MDa, $c/c^* = 0.14$ PEO solution ($\beta = 0.90$) using different DoS geometries, including nozzle radii $R_0$ and aspect ratios $\varLambda$ shown in \figref{fig:5}. The frame rate was $f_s= 4,000$ fps for all experiments.}
\begin{ruledtabular}
\begin{tabular}{cccccccccc}
$R_0$ (mm)  & $\varLambda$ & $Bo$ & $R^*$ (mm) & $t_R$ (ms) & $\alpha$ &  $\tau_E$ (ms) & $Oh$ $(\times 10^{-3})$ & $De$ & $Ec$ $(\times 10^{-4})$ \\
\midrule
		
0.359 & 2.5 & 0.10 & 0.07 & 0.88 & 0.57 & 2.88 & 7.38 & 3.27 & 2.19 \\
		
\midrule
		
\multirow{ 3}{*}{0.635} & 1.5 & 0.19 & 0.08 & \multirow{ 3}{*}{2.02} & 0.66 & 2.96 & \multirow{ 3}{*}{5.55} & 1.47 & 3.76 \\
 & 2.0 & 0.25 & 0.10 &  & 0.68 & 3.62 &  & 1.80 & 3.07 \\
 & 2.5 & 0.31 & 0.14 &  & 0.68 & 4.02 &  & 1.99 & 2.77 \\
		
\midrule
		
\multirow{ 3}{*}{1.055} & 1.0 & 0.35 & 0.08 & \multirow{ 3}{*}{4.31} & 0.45 & 2.51 & \multirow{ 3}{*}{4.31} & 0.58 & 7.37 \\
 & 1.5 & 0.52 & 0.11 &  & 0.59 & 2.69 &  & 0.62 & 6.88 \\
 & 2.0 & 0.69 & 0.15 &  & 0.63 & 3.30 &  & 0.76 & 5.61 \\
\end{tabular}
\end{ruledtabular}
\end{table*}

In \figref{fig:5} we show measurements of the minimum filament radius $R_{\mathrm{min}}(t)$ for a PEO solution with a molecular weight of $M_v = 8.1$ MDa, and a concentration $c/c^* = 0.14$. In \figref{fig:5}(\textit{a}) the nozzle radius is $R_0 = 0.635$ mm and the aspect ratio $\varLambda$, or nozzle height $H$ above the substrate, is manipulated such that different Bond numbers of $Bo = 0.19, 0.25$ and 0.31 are compared. The black dashed line in \figref{fig:5}(\textit{a}) represents a fit of the IC regime for a Bond number $Bo = 0.25$, yielding a pre-factor of $\alpha = 0.69$ and break-up time of $t_b = 3.3$ ms. The inset axis in \figref{fig:5}(\textit{a}) shows the dimensionless IC scaled filament radius, i.e., $(R_{\mathrm{min}}/R_0)^{3/2}$ versus $t/t_R$ such that the slope of \eqnref{eqn:1} is $-\alpha^{3/2}$. Within the margin of experimental uncertainty, values of the pre-factor $\alpha$ and break-up time $t_b$ for these experiments are not distinguishably different over this limited range of $Bo$. Fits of the EC regime using \eqnref{eqn:5} for the experiments at different Bond number also yield similar relaxation times that are approximately equal to $\tau_E = 3.5$ ms $\pm 0.5$ ms. However, the filament radius $R^*$ at the transition from IC to EC thinning varies systematically with $Bo$, and the values of $R^*$ are listed in \tabref{tab:4}. Consistent with the simulations of \citet{Zinelis2024}, \figref{fig:5}(\textit{a}) and \tabref{tab:3} show that the transition radius $R^*$ increases monotonically with Bond number $Bo$. 

\figref{fig:5}(\textit{b}) shows the evolution in the time-varying Weissenberg number $Wi(t)$ versus dimensionless time $(t-t^*)/\tau_E$ for the same PEO solution dripping from the $R_0 = 0.635$ mm nozzle at different Bond numbers $Bo$. All flows exhibit consistency with the theoretical Weissenberg number for EC thinning $Wi = 2/3 \pm 1/3$, with a visibly cylindrical filament, as shown by the inset axis in \figref{fig:5}(\textit{b}).

\figref{fig:5}(\textit{c}) shows DoS measurements for the same PEO solution from \figref{fig:5}(\textit{a}), but dispensed from a nozzle with a larger radius of $R_0 = 1.055$ mm. All flows shown in \figref{fig:5}(\textit{c}) with $R_0 = 1.055$ mm, have an Ohnesorge number of $Oh = 4.31\times10^{-3}$ (as listed in \tabref{tab:4}), smaller than the values shown in \figref{fig:5}(\textit{a}) with $R_0 = 0.635$ mm and $Oh = 5.55 \times 10^{-3}$. Furthermore, the nozzle height $H$ (or aspect ratio $\varLambda$) was also manipulated for the filament thinning experiments using the $R_0 = 1.055$ mm nozzle shown in \figref{fig:4}(\textit{c}), further manipulating the values of the Bond number attained. For the flow with the smallest Bond number ($Bo= 0.35$), inertio-capillary thinning is observed at early times with a pre-factor of $\alpha = 0.63$ and a break-up time of $t_b = 6.2$ ms,  similar to the results of \figref{fig:5}(\textit{a}) for flows with Bond numbers $Bo \leq 0.31$. However, for larger Bond numbers, free surface oscillations in the IC regime start to appear and fits using \eqnref{eqn:1} are more difficult to perform. However, once the EC regime is established all flows exhibit a similar extensional relaxation time $\tau_E = 2.9 \text{ ms} \pm 0.4 \text{ ms}$ based on fits using \eqnref{eqn:5}. Although not shown in \figref{fig:5}, measurements using the same fluid with the smallest nozzle radius of $R_0 = 0.359$ mm (and a Bond number of $Bo = 0.10$) were also performed, with the results listed in \tabref{tab:4}. Across all Bond and Ohnesorge numbers examined, the values of the extensional relaxation times $\tau_E$ agree well with one another, and the average extensional relaxation time was determined to be $\tau_E = 3.1 \text{ ms} \pm 0.5 \text{ ms}$. Lastly, \figref{fig:5}(\textit{d}) shows the evolution in the Weissenberg number $Wi(t)$ for the same fluid dispensed from the $R_0 = 1.055$ mm nozzle at different Bond numbers $Bo$. The inset profiles in \figref{fig:5}(\textit{d}) also demonstrate that in each DoS experiment a cylindrical filament is established near the middle of the EC regime, similar to the flows shown in \figref{fig:5}(\textit{b}) with a smaller nozzle radius.

\section{Discussion}

\subsection{Model-specific operability diagram}

\begin{figure*}
    \centering
    \includegraphics[width=\textwidth]{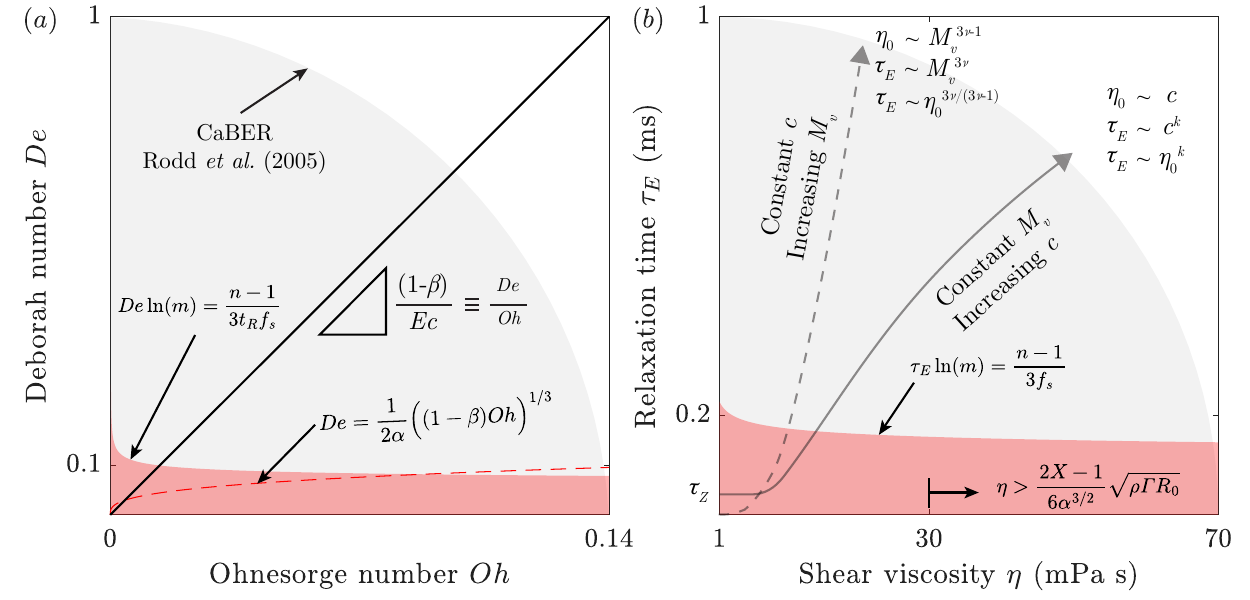}
    \caption{Nomograms for DoS operability shown on linear-linear plots of (\textit{a}) the Deborah number versus Ohnesorge number, and (\textit{b}) the extensional relaxation time versus viscosity. Within the grey shaded region, CaBER cannot be used to measure $\tau_E$ or $\eta$ according to \citet{Rodd2005}. The red shaded zones corresponds to the region in which a relaxation time $\tau_E$ cannot be reliably measured using DoS rheometry due to experimental limitations, i.e. \eqnref{eqn:16}, calculated assuming $n = 20$ data points and $f_s = 10,000$ fps. The red dashed line represents the theoretical limit of DoS rheometery based on \eqnref{eqn:12} assuming $\beta = 0.95$ and $\alpha = 0.64$. Both the experimental and theoretical limits correspond to a nozzle with a radius $R_0 = 0.635$ mm. For a Newtonian fluid, the viscosity can be reliably measured for Ohnesorge number $Oh>0.14$, or $\eta \gtrsim 30$ mPa s for DoS rheometry with a nozzle of radius $R_0 = 0.635$ mm, as annotated in (\textit{b}). The solid grey line in (\textit{b}) represents a typical trajectory through this operational state space for a dilute polymer solution at a fixed molecular weight $M_v$ and increasing concentration $c$. While the dashed line represents the corresponding trajectory for a dilute polymer solution of a given concentration $c$ with increasing molecular weight $M_v$.}
    \label{fig:6}
\end{figure*}

In the present study, limiting criteria have been derived for successfully measuring the dynamic viscosity $\eta$ (of a Newtonian fluid) and the extensional relaxation time $\tau_E$ (of a viscoelastic fluid) using DoS rheometry. Expressions for a minimum measurable extensional relaxation time (and corresponding minimum value of Deborah number) have been derived based on (a) scaling arguments for the thinning dynamics in in \S\S \ref{sec:2.A} and \ref{sec:2.B}, as well as (b) the spatial and temporal resolution of the measurement system in \S \ref{sec:2.C}. The limits on DoS operability are succinctly summarized in an operability diagram or \textit{nomogram} shown in red in \figref{fig:6}, that demarcates the regions in which the extensional relaxation time can be successfully measured. \citet{Rodd2005} developed the earliest version of such an operability diagram for capillary break-up extensional rheometry (CaBER). The grey shaded region in \figref{fig:6}(\textit{a}) depicts the values of the Deborah number and Ohnesorge number that cannot be reliably measured from CaBER, i.e., $De_{\mathrm{min}} =1$ and $Oh_{\mathrm{min}} = 0.14$, with similar dimensional representations (in terms of extensional relaxation time and viscosity) shown in \figref{fig:6}(\textit{b}). Overall, \figref{fig:6} demonstrates that DoS rheometry can achieve lower measurements of the extensional relaxation time compared to CaBER. This is primarily attributed to the smaller geometry and hence lower inertio-capillary time scale associated with DoS rheometery (i.e., the Rayleigh time $t_R$), as well as the better experimental resolution of modern high-speed imaging. The following section summarizes the nomograms for DoS extensional rheometry, starting with the theoretical limits that relate to the thinning dynamics.

Reliable measurements of the viscosity and extensional relaxation time using DoS rheometry depend on the Ohnesorge and Deborah number of the flow. Below a critical Ohnesorge number of $Oh \approx 0.14$, the filament thinning process is dominated by an inertio-capillary balance and measurements of the viscosity cannot be performed. For the relaxation time of a complex fluid to be reliably measured, the flow must be ``strong enough'' to trigger the establishment of an elasto-capillary balance, satisfying a Weissenberg number of $Wi(t^*) \geq 2/3$, where $t^*$ is the critical time at which the filament thinning dynamics transition from IC to EC thinning. When simplified, this critical value of the Weissenberg number (evaluated at the cross-over time to elastocapillary thinning) corresponds to a constraint on the Deborah number, represented by \eqnref{eqn:11} and detailed in \S\ref{sec:2.B}. 

Corresponding nomograms for DoS rheometry are also shown alongside that of CaBER in \figref{fig:6}. The red dashed line demonstrates the theoretical limit, according to \eqnref{eqn:11}, below which the Deborah number and extensional relaxation time cannot be reliably measured using DoS. We note that as the polymer concentration and molecular weight decreases, the viscosity ratio approaches unity $\beta \rightarrow 1$, and the theoretical minimum Deborah number approaches $De_{\mathrm{min}} \rightarrow 0$. In this case, however, measurements become increasingly limited by experimental constraints (i.e., the spatial and temporal resolution of the DoS setup), consistent with the conclusions of \citet{Sur2018}. The minimum Deborah number due to experimental limitations, including the sampling rate $f_s$ and spatial dynamic range $m$, is described by \eqnref{eqn:15} -- with equivalent representations provided in Appendix \ref{app:A} -- and this is represented by the red shaded region in \figref{fig:6}(\textit{a}).

\figref{fig:6}(\textit{b}) demonstrates the same operability diagram as \figref{fig:6}(\textit{a}), but in the dimensional space of extensional relaxation time $\tau_E$ versus dynamic viscosity $\eta$. The range of relaxation time $\tau_E$ that cannot be measured using DoS experiments are shown by the red shaded region, alongside the operability window for CaBER derived by \citet{Rodd2005} shown in grey. The limiting Ohnesorge number $Oh_{\mathrm{min}} = \eta/\sqrt{\rho \varGamma R_0} = 0.14$ translates to a minimum measurable viscosity $\eta_{\mathrm{min}}$ between 20 mPa s and 40 mPa s, depending on user selection of the nozzle radius $0.359 \text{ mm} <R_0 < 1.055 \text{ mm}$. For a nozzle radius of $R_0 = 0.635$ mm the minimum measurable viscosity is 30 \mbox{mPa s}, as annotated on \figref{fig:6}(\textit{b}) near the limit where $\tau_E \rightarrow 0$ ms. From the definition of $Oh$, the minimum measurable viscosity scales with the nozzle radius according to $\eta_{\mathrm{min}} \sim R_0^{1/2}$. On the other hand, experimental limitations on the range of $\tau_E$ only show a mild dependence on $R_0$, but are more greatly influenced by changes in the measurement parameters, such as the frame rate $f_s$, spatial dynamic range $m$, and the desired data point resolution $n$. Qualitatively, this explains why \citet{Sur2018} were able to measure extensional relaxation times of $\tau_E \sim \mathcal{O}(0.01)$ ms, given they used an exceptionally large sampling rate of $f_s = 50,000$ fps. The filament capture rate $\mathrm{FCR}$, described by \eqnref{eqn:16}, provides a metric that succinctly captures the relationship between the minimum measurable relaxation time and the spatio-temporal resolution of the DoS imaging setup. 

\begin{table*}
\caption{\label{tab:5}Comparison of the filament capture rate (FCR), other measurement parameters, and the minimum reported extensional relaxation time $(\tau_E)_{\mathrm{min}}$ for different experimental investigations that have employed DoS rheometry. Citations are listed in increasing order of the estimated values of FCR.}
\begin{ruledtabular}
\begin{tabular}{cccccccccc}
Citation & $f_s$ (fps) & $m$ & FCR (s$^{-1}$) & $(\tau_E)_{\mathrm{min}}$ (ms) & $n$ & $3\tau_E \cdot \mathrm{FCR}$ \\
\midrule
\citet{Soetrisno2023}  & 8,000 & 2.5 & 7,300 & 0.39 & 8 & 9 \\
\citet{Rosello2019}\footnote{The printer ink solutions used in \citet{Rosello2019} exhibit a transition from VC to EC thinning.} & 20,000 & 2.0 & 13,800 & 0.15 & 6 & 6 \\
\citet{Zhang2024} & 11,800 & 7.5 & 23,800 & 0.12 & 9 & 9 \\
Present work & 9,300 & 32 & 32,200 & 0.39 & 32 & 38 \\
\citet{Dinic2015}\footnote{Measurements of the minimum filament radius $R_{\mathrm{min}}(t)$ and $n$ are not shown in \citet{Dinic2015}; the estimate of $m \approx 10 $ is based on measurements for higher concentration PEO solutions shown in the same work.} & 25,000 & $\approx 10$ & 57,500 & 0.37 & - & 64 \\
\citet{Sur2018} & 50,000 & 3.4 & 61,100 & 0.02 & 5 & 4 \\
\end{tabular}
\end{ruledtabular}
\end{table*}

For a dilute polymer solution, navigating out of the immeasurable red-shaded region can be done in one of three ways.

\begin{enumerate} 
    \item For a constant molecular weight $M_v$, the concentration $c$ can be increased. In this case, the polymeric viscosity scales as $\eta_0 \sim c$ according to the first-order term of the Huggins \eqnref{eqn:26a}, and the extensional relaxation time is commonly found to vary as $\tau_E \sim c^k$, where $k$ is an exponent determined empirically to be $k \approx 0.67$ for PEO \citep{Dinic2015,Sur2018,Soetrisno2023}. An extensional relaxation time $\tau_E$ that varies with concentration contradicts simple dilute solution theories, such as the the Zimm model, which assumes the relaxation time is independent of polymer concentration $c$. It has been argued that this unexpected variation arises from weak self-concentration effects as the long polymer chains begin to unravel \citep{Prabhakar2016}. Careful measurements using DoS at sufficiently low concentrations do show an approach to a constant extensional relaxation time $\tau_E = \tau_Z$ that is independent of concentration $c$ \citep{Soetrisno2023}. This convergence of the measured relaxation time to the Zimm time $\tau_Z$ at low concentrations $c$ is annotated in the lower left corner of \figref{fig:6}(\textit{b}), where the solid grey line intersects the ordinate axis, and the viscosity of the solution approaches that of the solvent $\eta_0 \rightarrow \eta_s$ with decreasing $c$.
    
    \item For a constant concentration $c$, the molecular weight $M_v$ can be increased. In this case, dilute solution theory \citep{Larson1988} predicts that the polymer contribution to the viscosity will scale as $ \eta_p = (\eta_0 - \eta_s) \sim M_v^{3\nu-1}$ and the extensional relaxation time as $\tau_E \sim M_v^{3\nu}$, where $\nu$ is the solvent quality. The former scaling relies on the MHS and Huggins equations, i.e., $\eta_p \approx c[\eta]\eta_s$ and $\eta_0 \approx \eta_s(1+c[\eta])$. The change in relaxation time $\tau_E$ and viscosity $\eta_0$ with increasing molecular weight $M_v$ (for a constant concentration $c$) is represented by the dashed line in \figref{fig:6}(\textit{b}).
    
	\item Lastly, changing the solvent quality $\nu$ (for a given molecular weight $M_v$ and concentration $c$) will also change the relaxation time $\tau_E$ and viscosity $\eta_0$, based on the same scaling $\tau_E \sim M_v^{3 \nu}$ and $\eta_p \sim M_v^{3\nu-1}$. Dissolved salts and other ions in the solvent can greatly change the equilibrium polymer conformation and reduce the solvent quality $\nu$, hence decreasing the extensional relaxation time $\tau_E$ and viscosity $\eta_0$. 
\end{enumerate}

There are several options to capture additional constraints by displaying a third axis in the $De-Oh$ operability diagram shown in \figref{fig:6}(\textit{a}), for example the Bond number $Bo$. In \figref{fig:5}, we demonstrated that DoS rheometry of the same PEO solution at different Bond numbers (in the range $Bo < 1)$) do not demonstrate measurable differences in values of extensional relaxation time $\tau_E$. However, there are observed changes to the dynamics (e.g., the transition radius $R^*$) that can influence the criteria for the minimum measurable extensional relaxation time $\tau_E$.
As noted earlier, to assess the quality of a DoS experimental setup we have defined a ``figure of merit'' known as the \textit{filament capture rate} $\mathrm{FCR} = f_s\ln(m)$, where $f_s$ is the image acquisition rate (\textit{temporal resolution}) and $m$ is a value related to the image magnification (\textit{spatial resolution}) -- as defined in \S \ref{sec:2.C}. \tabref{tab:5} compares values of the FCR for a number of recent investigations that used DoS rheometry to measure relaxation times with $\tau_E \leq 1$ ms for a range of weakly elastic fluids -- including aqueous PEO \citep{Dinic2015,Sur2018} and PAM \citep{Soetrisno2023} solutions, printer inks \citep{Rosello2019}, and more complex systems, such as aqueous solutions of poly(\textit{N}-isopropylacrylamide) with the addition of a co-solvent \textit{N,N}-dimethylformamide \citep{Zhang2024}. The FCR is representative of the number of data points that are obtained in the elasto-capillary (EC) regime relative to the relaxation time, or $\mathrm{FCR} \approx (n-1)/(3\tau_E)$. The FCR is also a representation of the largest measurable extensional strain rate within the EC regime. Larger values of the FCR are generally required to reliably measure small values of the extensional relaxation time $\tau_E$, or enable better statistical accuracy by capturing a large number of data points ($n$) within the EC regime. To our knowledge, the smallest extensional relaxation time reported using DoS rheometry is $(\tau_E)_{\mathrm{min}} = 0.02$ ms, for an aqueous PEO solution with a molecular weight of $M_v = 0.2$ MDa and with a small nozzle of radius $R_0 = 0.400$ mm \citep{Sur2018}. The duration of the EC regime they measured was only $\Delta t = 0.08$~ms; however, an exceptionally high image acquisition rate of $f_s = 50,000$ fps enabled $n = 5$ images within the short duration of the EC regime, owing to the large value of the FCR (cf. \tabref{tab:5}).

The FCR can also be scaled, for example, by multiplication with an estimate for the relaxation time $\tau_E$. When normalized in such a way, the FCR can be used to predict the number of data points $n$ within the EC regime, i.e., $n \approx 3\tau_E \cdot \mathrm{FCR}$. This is useful if an experimental rheologist already has some \textit{a priori} estimate for the extensional relaxation time $\tau_E$ of the system (e.g., the Zimm time $\tau_Z$). \tabref{tab:5} lists the expected number of data points in the EC regime (i.e., $3\tau_E \cdot \mathrm{FCR}$), compared to the actual number of data points $n$ observed within the EC regime across these different experimental DoS investigations. Values of $n$ observed in the EC regime are nearly identical to those predicted using the FCR, demonstrating the robustness of the prediction based on the FCR. Therefore, recognizing the dependence of the FCR on the digital resolution $\mathcal{D}$ and frame rate, the spatial and temporal resolution of a DoS setup can be tuned to improve the FCR and achieve a larger number of data points ($n \gg 1$) within the EC regime. Improving measurement resolution can be done in one of two ways: first, by increasing the image acquisition rate $f_s$, and second by increasing the spatial dynamic range $m$. The former approach provides a more pronounced enhancement in resolution ($\mathrm{FCR} \sim f_s$) compared to the latter ($\mathrm{FCR} \sim \ln m$).

\subsection{Model-agnostic operability diagram}

\begin{figure*}
    \centering
    \includegraphics{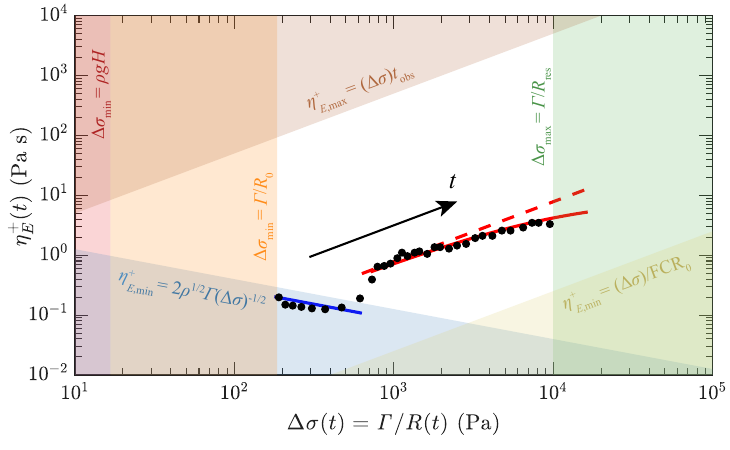}
    \caption{Model-agnostic operability diagram for DoS measurements of transient extensional viscosity $\eta_E^+(t)$ versus extensional stress difference $\Delta \sigma (t)$. The shaded coloured regions correspond to Eq. (\ref{eqn:20}),(\ref{eqn:21}), (\ref{eqn:22}), (\ref{eqn:23}), (\ref{eqn:24}), (\ref{eqn:25}), as annotated on the figure and previously defined in \S \ref{sec:2.D}. Within these regions, measurements of $\eta_E^+$ and $\Delta \sigma$ are not possible. Black symbols are the experimental measurements for the $M_v =  8.1$~MDa, $c/c^* = 0.06$, PEO solution shown in \figref{fig:1}. Here $R_0 = 0.359$ mm, $\varGamma = 63$ mN m$^{-1}$, $R_{\mathrm{res}} = 10$ \SI{}{\micro\meter}, the filament capture rate is $\mathrm{FCR}_0 = 36,000$ s$^{-1}$ (with $f_s = 10,000$ fps, and $m_0 = 36$), and $t_{\mathrm{obs}} = 10$ s (corresponding to an evaporation time scale with a mass transfer coefficient of $h_m \approx 10^{-5}$ m s$^{-1}$). The blue solid line represents IC thinning with $\alpha = 0.74$. The red dashed line represents EC thinning of a dilute solution of Hookean dumbbell with $\tau_E = 0.52$ ms. Lastly the red solid line represents the fitted trend in the transient extensional viscosity versus tensile stress difference for a nonlienar FENE-P dumbbell model ($L^2 = 90\times 10^3$) \citep{Wagner2015}.}
    \label{fig:7}
\end{figure*}

A limitation of the previously discussed guidelines is their narrow applicability to low-viscosity, viscoelastic fluids, such as dilute aqueous solutions of long-chain polymers. To ameliorate this concern we have developed model-agnostic limits in \S \ref{sec:2.D} for measuring the transient extensional viscosity $\eta_E^+(t)$ as a function of the capillarity-driven extensional stress difference $\Delta \sigma(t)$. These limits are encapsulated by the model-agnostic operability diagram of $\eta_E^+(t)$ versus $\Delta \sigma(t)$ shown in \figref{fig:7}, which can be constructed for any capillarity-driven extensional rheometry experiment, and any material (Newtonian or non-Newtonian).

In \S \ref{sec:2.D} and Appendix \ref{app:B} we have identified six limitations on measurements of $\eta_E^+(t)$ versus tensile stress difference $\Delta \sigma(t)$. Regions outside of these six limits are shown as shaded coloured regions in the state space of $\eta_E^+(t)$ versus $\Delta \sigma(t)$ presented in \figref{fig:7}. Measurements of the apparent transient extensional viscosity versus capillarity-driven stress difference for a representative PEO solution (with measurements of $R_{\mathrm{min}}(t)$ having been previously shown in \figref{fig:1}) are shown within these rheometry-limiting bounds. Two limits, corresponding to \eqnref{eqn:20} and \eqnref{eqn:21}, pertain to additional stresses (gravity and inertia) that can disrupt the desired balance between viscous (or elastic) stresses and the driving capillary stresses. For an extensional stress difference of $\Delta \sigma(t) \lesssim \rho g H$ (shown by the red shaded region in \figref{fig:7}), gravitational stresses can corrupt DoS measurements causing fluid to drain from the filament along the positive \textit{z-}direction. For extensional viscosities of $\eta_E^+(t) \lesssim 2\rho^{1/2}\varGamma(\Delta \sigma)^{-1/2}$, shown by the blue shaded region in \figref{fig:7}, the flow exhibits IC thinning with no measurable rheological parameter (be it viscosity or extensional relaxation time). Measurements of the PEO solution in \figref{fig:7} exhibit values of $\eta_E^+(t)$ that fall within the blue shaded region and a scaling of $\eta_E^+ \sim (\Delta \sigma)^{-1/2}$ (shown by the blue solid line), that is consistent with the observation of IC thinning at early times. At later times, the flow transitions to EC thinning with a transient extensional viscosity and stress difference that falls within the operability window. During the initial stages of EC thinning, the transient extensional viscosity exhibits a linear scaling with the stress difference, $\eta_E^+ \sim \Delta \sigma $, as derived in Appendix \ref{app:B}, and shown by the red dashed line in \figref{fig:7}. As the filament thins over time, and the tensile stress difference grows, polymer chains begin to reach their limit of finite extensibility, and deviate from the linear EC trend between $\eta_E^+$ and $\Delta \sigma$ predicted by analysis of the Hookean springs. At longer times and larger tensile stress differences, the transient extensional viscosity $\eta_E^+$ approaches a constant value, marking the limit of fully extended polymer chains. Fitting the measurements shown in \figref{fig:7} with the analytical model developed by \citet{Wagner2015} for a FENE-P fluid (shown by the red solid line in \figref{fig:7}) captures the effects of finite extensibility nicely. The FENE-P model deviates from the neo-Hookean limit at larger values of time and $\Delta \sigma \approx 2,000$ Pa, and exhibits a slow approach to a constant value of $\eta_E^+$ for $\Delta \sigma \gg 10^4$ Pa (albeit not detectable due to the resolution limit of the camera, $\Delta \sigma_{\mathrm{max}} = \varGamma/R_{\mathrm{res}}$).

The remaining four limitations pertain to experimental resolution constraints on capillarity-driven extensional rheometry. Two of these four constraints are caused by \textit{spatial resolution} and are defined by \eqnref{eqn:22} and \eqnref{eqn:23}. Measurements of the minimum filament radius are confined to a range of values between $R_{\mathrm{res}} \leq R_{\mathrm{min}}(t) \leq R_0$; therefore, the extensional stress difference, defined by \eqnref{eqn:18}, is equally constrained to $\varGamma/R_0 \leq \Delta \sigma (t) \leq \varGamma/R_{\mathrm{res}}$ and represented by the orange and green shaded regions in \figref{fig:7}. Measured values of $\Delta \sigma$, shown with black symbols in \figref{fig:7}, are contained within the region between these two limits, where the ratio of the two extremes on $R_{\mathrm{min}}$, or $\Delta \sigma$, is the total spatial dynamic range $m_0 = \Delta \sigma_{\mathrm{max}}/ \Delta\sigma_{\mathrm{min}} = R_0/R_{\mathrm{res}}$. Taking a ratio of the analytical minimum stress difference caused by gravitational stresses, $\Delta \sigma_{\mathrm{min}} = \rho g H$, to the experimental minimum stress difference, $\Delta \sigma_{\mathrm{min}} = \varGamma/R_0$, recovers the Bond number $Bo = \rho g H R_0/\varGamma$, which is generally less than unity for most flows considered.

Lastly, limits on \textit{temporal resolution}, described by \eqnref{eqn:24} and \eqnref{eqn:25}, are represented by the yellow and brown shaded regions in \figref{fig:7}. As detailed in Appendix \ref{app:B}, constraints on the temporal resolution confine measurements of the transient extensional viscosity to a range of values between $(\Delta \sigma)/\mathrm{FCR}_0 \leq \eta_E^+(t) \leq (\Delta \sigma) \cdot t_{\mathrm{obs}}$. The maximum criteria $\eta_{E,\mathrm{max}}^+ = (\Delta \sigma) \cdot t_{\mathrm{obs}}$ corresponds to a limiting observational time $t_{\mathrm{obs}}$, attributed to (for example) drying, evaporation or sample solidification. The minimum experimental criterion for the extensional viscosity $\eta_{E,\mathrm{min}} = (\Delta \sigma)/\mathrm{FCR}_0$ corresponds to the largest measurable extensional strain rate, where $\mathrm{FCR}_0 = f_s \ln(m_0)$ is a model-agnostic version of the \textit{filament capture rate} that is based on the total spatial dynamic range $m_0$. Similar to the model-specific criteria, i.e., the spatial dynamic range $m$ and filament capture rate $\mathrm{FCR}$, the model agnostic filament capture rate can be improved by augmenting the sampling rate $f_s$ or the total spatial dynamic range $m_0$ -- for example by increasing the imaging magnification and reducing $R_{\mathrm{res}}$.

Collectively, these six model-agnostic criteria bound an operability window (shown in white in \figref{fig:7}) that can be tuned and widened for measuring the transient extensional viscosity of liquid-like materials. For a typical DoS setup with parameters similar to those depicted in \figref{fig:7}, the transient extensional viscosity can be measured for fluids with an extensional viscosity given by $\eta_E^+ \gtrsim 0.1$ Pa s. A very similar model-agnostic operability diagram can be constructed for other capillarity-driven thinning rheometers such as CaBER; the principal quantitative difference between the operability windows is the value of $R_0$ used in each device, which is commonly $R_0 = 6$~mm for CaBER and $R_0 \sim O(1)$~mm for a DoS instrument. 

Viewed holistically, these guidelines help experimental rheologists maximize the resolution of their measurements and provide bounds on the measured rheological parameters -- in terms of both a model-agnostic property, the (apparent) transient extensional viscosity $\eta_E^+$ as a function of an imposed tensile stress difference $\Delta \sigma$, as well as in terms of model-specific ($\eta$, $\tau_E$) measures. These simple definitions will help provide insights and simple physical scalings for the extensional material properties of a range of weakly elastic fluid materials such as dilute solutions of polymers, wormlike micelles, and rigid rods, that are difficult to measure in other test configurations.

\section*{Declarations}

The authors report no conflict of interest.

\section*{Acknowledgments}

The first author acknowledges the support of the Natural Sciences and Engineering Research Council of Canada (NSERC), [PDF-587339-2024]. Cette recherche a \'{e}t\'{e} financ\'{e}e par le Conseil de recherches en sciences naturelles et en g\'{e}nie du Canada (CRSNG), [PDF-587339-2024].

\section*{Data Availability}

The data that support the findings of this study are available from the corresponding author upon reasonable request.

\appendix

\section{An explicit criteria for determining the minimum measurable relaxation time}\label{app:A}

The following appendix provides an explicit form of the inequality (\ref{eqn:16}) that can be used to determine the minimum measurable relaxation time of a DoS measurement based on experimental constraints. As defined in \S \ref{sec:2.C}, the spatial dynamic range of a DoS measurement depends on the transition from IC to EC thinning at a critical radius $R^*$. This gives rise to a spatial dynamic range $m = R^*/R_{\mathrm{res}}$, where $R_{\mathrm{res}}$ is the minimum measurable radius based on the constraints of the imaging magnification. Incorporating \eqnref{eqn:6} for $R^*$ into the inequality (\ref{eqn:16}) yields

$$	
\tau_E \ln \Big(\frac{Ec}{2} m_0^3 \Big) \geq \Delta t
$$

\noindent
where $\Delta t  = (n-1)/f_s$, and $m_0 = R_0/R_{\mathrm{res}}$ is the total spatial dynamic range (based on the nozzle radius $R_0$). Given the elasto-capillary number is $Ec = G R_0/\varGamma$, where $G = \eta_p / \tau_E$, the above inequality can also be written as

$$
\tau_E \ln \Big(\frac{\eta_p R_0}{2\tau_E\varGamma}m_0^3\Big) \geq \Delta t.
$$

\noindent
This implicit inequality for the range of extensional relaxation times $\tau_E$ that can be measured can also be expressed explicitly as

\begin{subequations}
	\begin{align}
	\tau_E & \geq t_{\mathrm{DoS}} \exp\Big[ \mathcal{W}_{-1}\Big(-\frac{\Delta t}{t_{\mathrm{DoS}}}\Big) \Big], \text{ } \tau_E \in (0,t_{\mathrm{DoS}}/e), \label{eqn:A1a}\\
	\text{where} \nonumber \\
	t_{\mathrm{DoS}} & = \frac{\eta_pR_0}{2\varGamma}m_0^3 = \frac{1}{2}(1-\beta)m_0^3\text{ }t_{\nu},
	\label{eqn:A1b}
	\end{align}
\end{subequations}

\noindent
is a time constant that contains the viscocapillary time scale of the fluid and the properties of the imaging system, while $\mathcal{W}$ is the Lambert function. The Lambert function is the inverse of the solution to the implicit function $w \exp(w)=z$, i.e., $w = \mathcal{W}(z)$, where $w$ and $z$ are arbitrary variables. Here the $-1$ branch of the Lambert function (i.e., $\mathcal{W}_{-1}$) is used, provided the DoS time scale $t_{\mathrm{DoS}}$ satisfies $t_{\mathrm{DoS}}/\Delta t>e$; a condition that is almost always true given $m_0 \sim \mathcal{O}(10^{2})$ and $\Delta t \sim \mathcal{O}(10^{-3})$~s. A similar explicit constraint for the Deborah number can be expressed by dividing \eqnref{eqn:A1a} by the Rayleigh time to obtain

\begin{subequations}
	\begin{align}
	De & \geq C \exp\Big[ \mathcal{W}_{-1}\Big(-\frac{\Delta t}{t_R C}\Big)\Big],  \text{ } De \in (0,C/e) \label{eqn:A2a}\\
	\text{where} \nonumber \\
	C & =  \frac{(1-\beta)Oh}{2}m_0^3 = \frac{t_{\mathrm{DoS}}}{t_R},
	\label{eqn:A2b}
	\end{align}
\end{subequations}

\noindent
is a dimensionless representation of the DoS time constant, scaled using the Rayleigh timescale of the fluid.

\section{Model agnostic operating limit derivations}\label{app:B}

In \S \ref{sec:2.D}, measurement limitations of the transient extensional viscosity $\eta_E^+(t)$ and capillarity-driven extensional stress difference $\Delta \sigma(t)$ are listed. We identify six limits, and the following appendix provides derivations and additional information on the physical interpretations of each constraint. The limits are detailed in the same order they are listed in \S \ref{sec:2.D}. We begin with limits pertaining to gravitational and inertial stresses.

\subsection{Gravitational and inertial stresses}

Measuring the transient extensional viscosity is predicated on establishing a balance of between extensional viscous stress ($\eta_E^+ \dot{\varepsilon}$) and a capillary stress ($\Delta \sigma = \varGamma/R_{\mathrm{min}}$), as defined in \eqnref{eqn:19}. The 1-D model of \eqnref{eqn:19} assumes curvature along the axial $z-$direction is negligible and there are no additional contributions to the tensile stress difference arising in the thinning fluid filament, for example due to gravity or inertia. 

We begin by considering the influence of gravitational stresses, which can cause liquid to drain from the filament along the positive $z-$direction. To assume there is no gravity-driven drainage in the liquid filament, the ratio of gravitational to viscous stresses must be negligible. Gravitational stresses scale approximately with the height of the droplet $H$ along the axial $z-$direction. Hence, if gravitational drainage from the liquid filament is negligible, the tensile stress difference must be greater than gravitational stress $\Delta \sigma \gtrsim \rho g H$, as defined by \eqnref{eqn:20} and shown by the red shaded region in \figref{fig:7}.

In addition to considering the stresses in the filament, we also consider a ratio of time scales -- similar to the analysis done in \S \ref{sec:2.A} and \ref{sec:2.B}. The characteristic time scale $t_e$ for fluid to be drained from the filament depends on the instantaneous viscosity in the fluids and can be written as

$$t_e = \frac{\Delta \sigma}{\eta_E^+}.$$
\label{eqn:viscoelastic timescale}
\noindent
A similar approach can be taken to estimate the imposition of inertial stresses. Commonly seen in the unsteady Bernoulli equation, e.g., see \citet{Day1998}, inertial stresses scale according to

$$\rho \Big(\frac{\partial \phi}{\partial t} + \frac{1}{2}|\nabla\phi|^2 \Big) \sim \rho\frac{R_{\mathrm{min}}^2}{t_i^2},$$

\noindent
where $\boldsymbol{v} = \nabla \phi$ is the velocity vector, $\phi$ is the velocity potential (with dimensions of $[\mathrm{L}]^2[\mathrm{T}]^{-2}$) and $t_i$ is a representative inertial time scale. Balancing this inertial stress with the capillary stress $\Delta \sigma$, the inertial time scale can be represented as

$$t_i = \frac{\rho^{1/2} R_{\mathrm{min}}}{(\Delta \sigma)^{1/2}} = \frac{\rho^{1/2}\varGamma}{(\Delta \sigma)^{3/2}},$$
\label{eqn:inertial timescale}
\noindent

where the second expression can be obtained by substituting $R_{\mathrm{min}} = \varGamma/(\Delta \sigma)$ so that the expression is a function of the variable plotted on the abscissa of \figref{fig:7}. For inertial stresses to be negligible compared to the tensile stress difference in the filament, the time scale associated with extensional viscosity $t_e$ must be greater than the time scale for inertial breakup. Taking this ratio we obtain

$$\frac{t_e}{t_i} = \frac{\eta_E^+ (\Delta \sigma)^{1/2}}{\rho^{1/2}\varGamma} \geq \mathcal{O}(1).$$

\noindent
Rearranging the above inequality to be in terms of the transient extensional viscosity (assuming the critical ratio is unity), we establish a  criterion to neglect inertial resistance to capillary thinning when  
\begin{equation}
    \eta_E^+(t) \gtrsim \rho^{1/2}\varGamma(\Delta \sigma)^{-1/2}.
    \label{eqn:B1}
\end{equation}

\noindent
The above inequality approximately describes the trend in the ``pseudo'' apparent transient extensional viscosity $\eta_E^+$ versus stress difference $\Delta \sigma$ for IC thinning. We refer to this as a pseudo transient extensional viscosity because IC thinning has no real physical connection to a measurable rheological feature of the fluid, e.g. shear viscosity or extensional relaxation time. That said, using the definition of the transient extensional viscosity \eqnref{eqn:19}, this pseudo transient extensional viscosity for IC thinning can be directly established,

\begin{equation}
    (\eta_E^+)_{\mathrm{ic}} = \frac{3}{4}\frac{\rho^{1/3}\varGamma^{2/3}}{\alpha}(t_b-t)^{1/3} = \frac{3}{4\alpha^{3/2}}\rho^{1/2}\varGamma(\Delta \sigma)_{\mathrm{ic}}^{-1/2}.
    \label{eqn:B2}
\end{equation}

\noindent
Where the tensile stress difference in the IC regime $(\Delta \sigma)_{\mathrm{ic}}$ can be determined by substituting \eqnref{eqn:1} for $R_{\mathrm{min}}(t)$ of IC thinning in \eqnref{eqn:18}. Comparing \eqnref{eqn:B2} with \eqnref{eqn:B1} derived from scaling arguments, both exhibit scaling of $\eta_E^+ \sim (\Delta \sigma)^{-1/2}$; however, \eqnref{eqn:B1} is lacking a pre-factor equal to $3/(4\alpha^{3/2})$. \citet{Day1998} demonstrated that $\alpha = 0.64$, hence this prefactor amounts to $3/(4\alpha^{3/2}) = 1.5$. If the effects of inertia are to be considered negligible, the transient extensional viscosity must be greater than the pseudo transient extensional viscosity for IC thinning,

\begin{equation}
    \eta_E^+(t) \geq \frac{3}{4\alpha^{3/2}} \rho^{1/2}\varGamma \Big(\Delta \sigma(t) \Big)^{-1/2} \approx 2 \rho^{1/2}\varGamma \Big(\Delta \sigma(t) \Big)^{-1/2},
    \label{eqn:B3}
\end{equation}

\noindent
where in the definition provided by \eqnref{eqn:21} of \S \ref{sec:2.D}, we conservatively assume the pre-factor is $3/(4\alpha^{3/2}) \approx 2$, corresponding to $\alpha \approx 0.5$, which is on the lower range of values for $\alpha$ reported in the literature \citep{Dinic2015,Mckinley2005,Day1998}. Although this pre-factor might seem insignificant, if it was not included and \eqnref{eqn:B1} were used instead of \eqnref{eqn:B3}, the criterion would not include the IC regime of the measurements shown in \figref{fig:7}. This is because for the measurements of PEO shown in \figref{fig:7}, $\alpha = 0.74 > 0.5$; therefore, the blue shaded region spans all IC flows (with a similar density and surface tension to water) with $\alpha \geq 0.5$. Note, that for viscocapillary thinning of a Newtonian fluid the transient extensional viscosity is constant with respect to time and equal to

$$(\eta_E^+)_{\mathrm{vc}} = \frac{3}{2X-1}\eta.$$

\noindent
For EC thinning, the transient extensional viscosity increases linearly with respect to the capillarity-driven stress difference,

$$ (\eta_E^+)_{\mathrm{ec}} = \frac{3\tau_E}{2}(\Delta \sigma)_{\mathrm{ec}},$$

\noindent
where the tensile stress difference in the EC regime $(\Delta \sigma)_{\mathrm{ec}}$ can be determined from \eqnref{eqn:5} and \eqnref{eqn:18}.

\subsection{Experimental Resolution Limits}

The experimental resolution limits for DoS measurements can be separated into constraints based on \textit{spatial resolution} and \textit{temporal resolution}. The limits due to spatial resolution are simple to define; the minimum filament radius can only be measured for $R_{\mathrm{res}} \leq R_{\mathrm{min}} \leq R_0$, and therefore, the  capillarity-driven stress difference is confined to values between $\varGamma/R_0 \leq \Delta \sigma \leq \varGamma/R_{\mathrm{res}}$, as shown in \S \ref{sec:2.D} and defined by \eqnref{eqn:22} and \eqnref{eqn:23}. For adequate temporal resolution of a DoS measurement, we assert that the extensional strain rate $\dot{\varepsilon}(t)$, given by \eqnref{eqn:8}, must be suitably measured over time. To estimate the maximum extensional strain rate of a typical DoS measurement we approximate the extensional strain rate as

$$\dot{\varepsilon} = -2 \frac{d \ln (R_{\mathrm{min}})}{dt} \sim -\frac{\delta  \ln(R_{\mathrm{min}})}{\delta t},$$

\noindent
where $\delta$ is used to represent a finite incremental change in $\ln(R_{\mathrm{min}})$ and $t$. The largest (negative) incremental change in $  \ln(R_{\mathrm{min}})$ that can be imaged is

$$\delta \ln(R_{\mathrm{min}}) = \ln(R_{\mathrm{res}})-\ln(R_0) = -\ln(m_0),$$

\noindent
while the smallest incremental change in time is $\delta t = f_s^{-1}$. Therefore, a DoS measurement is confined to extensional strain rates of

$$\dot{\varepsilon} \leq f_s \ln(m_0) \equiv \mathrm{FCR}_0$$

\noindent
where $\mathrm{FCR_0}$ is the \textit{filament capture rate} based on the total spatial dynamic range $m_0 = R_0/R_{\mathrm{res}}$. To establish the minimum extensional strain rate we consider the incremental change in $\ln(R_{\mathrm{min}})$ to be relatively small, such that $\delta \ln(R_{\mathrm{min}}) = -1$, in comparison to a large incremental change in time $\delta t = t_{\mathrm{obs}}$, where $t_{\mathrm{obs}} \gg f_s^{-1}$ is the longest feasible observation time, that is set by extrinsic factors, such as evaporation, drying, or solidification, as discussed in \S \ref{sec:2.D}. Overall, the extensional strain rate is therefore constrained between a range of values $t_{\mathrm{obs}}^{-1} \leq \dot{\varepsilon} \leq \mathrm{FCR}_0$. Given the definition of the transient extensional viscosity \eqnref{eqn:19} in terms of a tensile stress difference and a strain rate, these constraints on the range of extensional strain rates correspond to limits on the extensional viscosity of $(\Delta \sigma)/\mathrm{FCR}_0 \leq \eta_E^+ \leq (\Delta \sigma)t_{\mathrm{obs}}$, as defined by \eqnref{eqn:24} and \eqnref{eqn:25}.

\section*{References}
\bibliography{bibliography}

\end{document}